\documentclass[a4paper, amsfonts, amssymb, amsmath, reprint, showkeys, nofootinbib]{revtex4-1}
\usepackage[english]{babel}
\usepackage[utf8]{inputenc}
\usepackage[colorinlistoftodos, color=green!40, prependcaption]{todonotes}
\usepackage[pdftex, pdftitle={Article}, pdfauthor={Author}]{hyperref}
\usepackage{float}
\usepackage{graphicx}
\usepackage{yhmath}
\usepackage{subfigure}
\usepackage{mathdots}
\usepackage{MnSymbol}

\begin{document}
\title{Quantized Vortices Mediated Annealing }
	\author{Airat Kamaletdinov$^1$ and Natalia G. Berloff$^{1,2}$ }
	\email[correspondence address: ]{N.G.Berloff@damtp.cam.ac.uk}
	\affiliation{$^1$Skolkovo Institute of Science and Technology, Bolshoy Boulevard 30, bld.1,
		Moscow, 121205, Russian Federation}
	\affiliation{$^2$Department of Applied Mathematics and Theoretical Physics, University of Cambridge, Cambridge CB3 0WA, United Kingdom}
	
	\begin{abstract}{The complex matter-field lattice model is a ubiquitous and universal physics model that directly links to many universal spin models. However, finding the ground state of such a model for the most general interactions between the lattice sites is an NP-hard problem. The knowledge of the ground state is crucially essential for the range of applications from quantum annealing to predicting new materials and understanding the behaviour of spin glasses. Here we show how letting the quantised quasi-3D vortex loops pass through the system during the dissipative quench dynamics helps it find its ground state. Such quantized vortices mediated annealing (QVMA) is achieved by introducing and dynamically changing selected couplings between the lattice sites that either affect the local dimensionality of the medium or change the local mobility of the defects. Analysing the performance of QVMA on strongly glassy systems, we demonstrate that QVMA represents a highly efficient mode of physics enhanced computation.  }
	\end{abstract}
	
	\maketitle
	{\it Introduction.}
    Many different physical implementations and models for quantum and classical analog computing  have been proposed in recent years. The physical implementations include superconducting qubits, neutral atoms in optical lattices, trapped ions, nuclear magnetic resonances, quantum dots, photons, nuclear spins, optical computing systems \cite{georgescu2014quantum}.  Each physical implementation assumes a particular model of computation ranging from the quantum circuit model characteristic of a universal quantum computation \cite{nielsen2002quantum}, adiabatic quantum annealing \cite{farhi2001quantum}, topological computing based on the braiding of a certain type of quasiparticles called anyons \cite{wilczek1982quantum}, quantum random walk \cite{aharonov1993quantum}, quantum Turing machines \cite{bernstein1997quantum}, measurement based quantum computing, permutational computing \cite{jordan2010quantum}, the one clean qubit model \cite{moussa2010testing}, gain-dissipative optical systems \cite{marandi2014network, nixon2013observing,berloff2017realizing, memristors}. The variability of models of quantum computation provides  alternative frameworks for both discovering new quantum algorithms and devising new physical implementations of quantum computers or simulators. In particular, the adiabatic model seems particularly promising for implementation in superconducting systems \cite{kaminsky2004scalable}, while loss minimisation of gain-dissipative systems is suitable for optical implementation \cite{marandi2014network,nixon2013observing,berloff2017realizing}.
    Simulators can be classified into digital when they use discrete quantum-gate operations and analog when they implement a spin Hamiltonian in an analog fashion \cite{buluta2009quantum}. The former is hugely demanding technology to design and build. It requires single quantum particles to be localized and controlled for long enough to apply hundreds or thousands of precise manipulations to them and then measured individually with high fidelity. The latter type of simulators is considered the most promising type of simulators achievable on the shorter time scale using a wide variety of experimental platforms. A particular type of computationally intractable problem that such simulator can address consists of finding a global minimum of universal classical spin models such as Ising, XY, Potts, etc., with the most general coupling matrix.  The universality means that any minimization problem of interest can be mapped into any of these  universal classical spin models with at most polynomial overhead on the number of variables \cite{de2016simple}. For instance, the minimization of spin Hamiltonians is performed in quantum annealing machines using adiabatic mode of quantum computation (e.g. D-wave) and in optical computing using the gain-dissipative principle of computation (e.g. Coherent Ising Machines).  Such {\it physics enhanced computation} is an emergent field of research that promises to facilitate the developments in machine learning, neural networks and large-scale optimization.

    In this Letter, we propose and justify a novel model of physics enhanced computation based on quantized vortices that  aims at finding the ground state of the Hamiltonian of the universal complex matter field lattice model \cite{svistunov2015superfluid}
    \begin{equation}
    H=-\frac{1}{2}\sum_{i,j}J_{ij}(\psi_i^*\psi_j+c.c) + \frac{g}{2}\sum_i(1-|\psi_i|^2)^2,
    \label{hpsi}
    \end{equation}
 where $\psi_i$ represents the wavefunction associated with the $i-$th lattice site,  $g>0$ is the strength of the nonlinear repulsive self-interactions  within the $i-$th site, and $J_{ij}$ is the strength of the coupling between the lattice sites $i$ and $j$. A large variety of the classical and quantum complex-field systems on a lattice from ultra-cold BECs in an optical lattice to superconducting qubits, planar magnets,
2D solids, Josephson-junction arrays, photon and polariton systems, superfluid and superconductor films, etc.  realise this Hamiltonian   \cite{domb2000phase}. The correspondence  of this model with the XY model can be seen by observing that the second term on the right-hand side represents the penalty for the deviation of the occupation $|\psi_i|^2$ from one. For sufficiently strong nonlinear self-interactions, therefore, the ground state (GS) configuration favours $\psi_i=\exp(i \theta_i)$ and reduces to the XY model $H=H_{XY}.$ 
     The XY  Hamiltonian  for $N$ spins denoted as a vector on a unit circle ${\bf s}_i=(\cos \phi_i, \sin \phi_i)$  takes form
   $
    H_{XY}=-\frac{1}{2}\sum_{i=1}^N\sum_{j=1}^N J_{ij}{\bf s}_i\cdot{\bf s_j} =-\frac{1}{2}\sum_{i=1}^N \sum_{j=1}^N  J_{ij}\cos (\phi_i-\phi_j).
   $

The time evolution of the system obeys the Hamiltonian dynamics $i \dot{\psi_i}=\partial H/\partial \psi_i^*$ and for Eq.~(\ref{hpsi}) becomes
    \begin{equation}
        i\frac{d \psi_i}{d t} = -  \sum_{j,j\ne i} J_{ij} \psi_j + g( 1 - |\psi_i|^2) \psi_i.
        \label{GP}
       \end{equation}
    If the particular case of $J_{ij}=J>0$  representing the nearest neighbor (NN) interaction on a square grid, Eq.~(\ref{GP}) reduces to  the discrete Gross-Pitaevskii model \cite{kevrekidis2001discrete} where the Laplacian operator  is replaced with its discretized version. Low-density Bose–Einstein condensates (BECs) embedded into optical lattices also obey a lattice model with $J_{ij}=J$ (NN interactions of the Josephson type) when  the tight-binding approximation is employed \cite{ brazhnyi2004theory}. The sign and magnitude of the interactions can be engineered to take a range of values \cite{struck2011quantum}  and beyond NN \cite{davis2019photon}. From the point of global minimisation, we are interested in the spin-glass couplings that include long-range interactions (or even all-to-all) so we consider the  sign and  magnitude varying  $J_{ij}$. The systems governed by anisotropic and  potentially frustrating couplings are ubiquitous from magnetic nanoparticle ensembles to molecular clusters bonded by electric dipolar and quadrupolar interactions \cite{gallina2020disorder}. Finding GS of these systems is a highly non-trivial task  and sophisticated algorithms borrowed from computer science  are often employed. 
    
    
    In this Letter, we address the question of   bringing the system to the ground state by experimentally accessible means and, therefore, to the GS of the lattice model having two crucial applications in mind.

    Firstly, the quantum annealing protocol  starts with preparing the system in a GS of an "easy" (often referred to as the "driver") Hamiltonian \cite{kadowaki1998quantum, farhi2001quantum}. The ground state of the driver Hamiltonian is assumed to be unique and easily prepared. The drivers are routinely taken as the ground state of the system  with ferromagnetic couplings $J_{ij}=J>0$ \cite{hen2016driver}. However, the connectivity of the driver Hamiltonian is not arbitrary. The embedding of practical optimization problems on experimental quantum annealers requires the construction of driver Hamiltonians that commute with the constraints of the problem \cite{hen2016quantum}. The resulting connectivity (adjacency) matrix is likely to contain lower-dimensional subsystems (1D and 2D), preventing the GS achievement of the driver Hamiltonian. We show that the nonlinearity, dissipation and quasi-3D nonplanar adiabatic connections can resolve this problem while driving the system to the true GS. 
    
    Secondly and more importantly, we show that the analysis of local mobility of quantised vortices  in the physical system can be used  to develop an annealing scheme (that does not have to be adiabatic)  to drive the spin glass to a lower energy state and even to GS with  dramatically  improved  success probabilities in comparison with  unassisted quenched dynamics or with annealing from the ferromagnetic state.

    
	{\it Classical lattice model meets Khalatnikov's two-fluid model.} Equation (\ref{GP}) describes  the time dynamics of a conservative Hamiltonian system:  a condensate, however,  has no preferred reference frame. In the actual experiments, the reference frame is set by noncondensed or reservoir particles that have to enter the theory.  For instance,  Khalatnikov’s  theory of mutual friction between the superfluid and normal fluid at finite temperatures involves three independent kinetic coefficients to describe the energy transfer between the components \cite{khalatnikov1965introduction}. According to this theory,  the integrated superfluid momentum equation is 
    $
        \phi_t + \frac{1}{2} u^2 +\mu = -\zeta_3 \nabla \cdot(\rho {\bf w}),
    $
   where $\phi$ is the phase of the field, ${\bf u}=\nabla\phi$ is the superfluid velocity, 
  $\rho$ is the superfluid density, $\mu$ is the chemical potential, ${\bf w}$ is the velocity of the normal fluid relative to the superfluid and $\zeta_3>0$ is the third kinetic coefficient of Khalatnikov's theory. Using the  equation of mass conservation $\rho_t+\nabla \cdot (\rho {\bf u})=0$  and assuming that the normal component is stationary in our reference frame (${\bf w}=-{\bf u}$) we obtain 
  \begin{equation}
  \frac{\partial \phi}{\partial t} + \frac{1}{2} u^2 +\mu = -\zeta_3 \frac{\partial \rho}{\partial t}.
  \label{khal2}
  \end{equation}
    To modify Eq.~(\ref{GP}) according to Khalatnikov's theory, we need to consider instead
  \begin{equation}
    \begin{aligned}
        i\frac{d \psi_i}{d t} =  &-  \sum_{j,j\ne i} J_{ij} \psi_j + g\;( 1 - |\psi_i|^2 ) \psi_i \\
        &-   i K \psi_i\sum_{j,j\ne i} J_{ij} ( \psi_j^* \psi_i - \psi_j \psi_i^* ),
        \label{main}
    \end{aligned}
 \end{equation}
 where $K=\zeta_3$. To see  the correspondence with Eq.~(\ref{khal2})  we use the Madelung transformation $\psi_i=\sqrt{\rho_i}\exp[i \phi_i]$ in Eq.~(\ref{main}) so that  the imaginary part yields 
    $
\dot{\rho_i}=2 \sum J_{ij} \sqrt{\rho_i\rho_j}\sin\phi_{ij}$ and the real part becomes $
\dot{\phi_i}-\sum J_{ij}\sqrt{\frac{\rho_j}{\rho_i}}\cos \phi_{ij}+g\;(1-\rho_i) =-K \dot{ \rho_i},
$
  where we denoted $\phi_{ij}=\phi_i-\phi_j$. If the continuous wavefunction is discretized on a square grid with the lattice constant  $\Delta$ and $J_{ij}=1/2\Delta^2$ NN interactions,  then $-\sum J_{ij}\cos \phi_{ij} \approx - \frac{1}{2\Delta^2}(4- \frac{1}{2}(\phi_i- \phi_{i\pm1})^2-\frac{1}{2}(\phi_i-\phi_{i\pm N})^2)$ $ \approx const +\frac{1}{2}(\nabla \phi)^2\approx const + \frac{1}{2} u^2.$  Associating $\mu = const +g(1-\rho_i)$ and $\rho_i \approx \rho_j$ completes the justification  of the Khalatnikov's correction to Eq.~(\ref{GP}).  We note that Eq.~(\ref{main}) for $J_{ij}=J>0$ and NN interactions only (and the time-dependent $K$) were used to study the non-equilibrium quenches \cite{de2013lyapunov,tarkhov2020} in laser-induced melting of the charge-density-wave order in solid \cite{zong2019evidence}.  The steady state is achieved  when $\sum J_{ij} \sqrt{\rho_i\rho_j}\sin\phi_{ij}=0$ which corresponds to the minimum of the XY model if $\rho_i=1$ as the penalty in the Hamiltonian $H$ dictates.
  Now we consider the quenching dynamics of the system described by Eq.~\ref{main} and determine the conditions that prevent the system from achieving its GS. 
  
\begin{figure}[t]
        \centering
        \includegraphics[width=\linewidth]{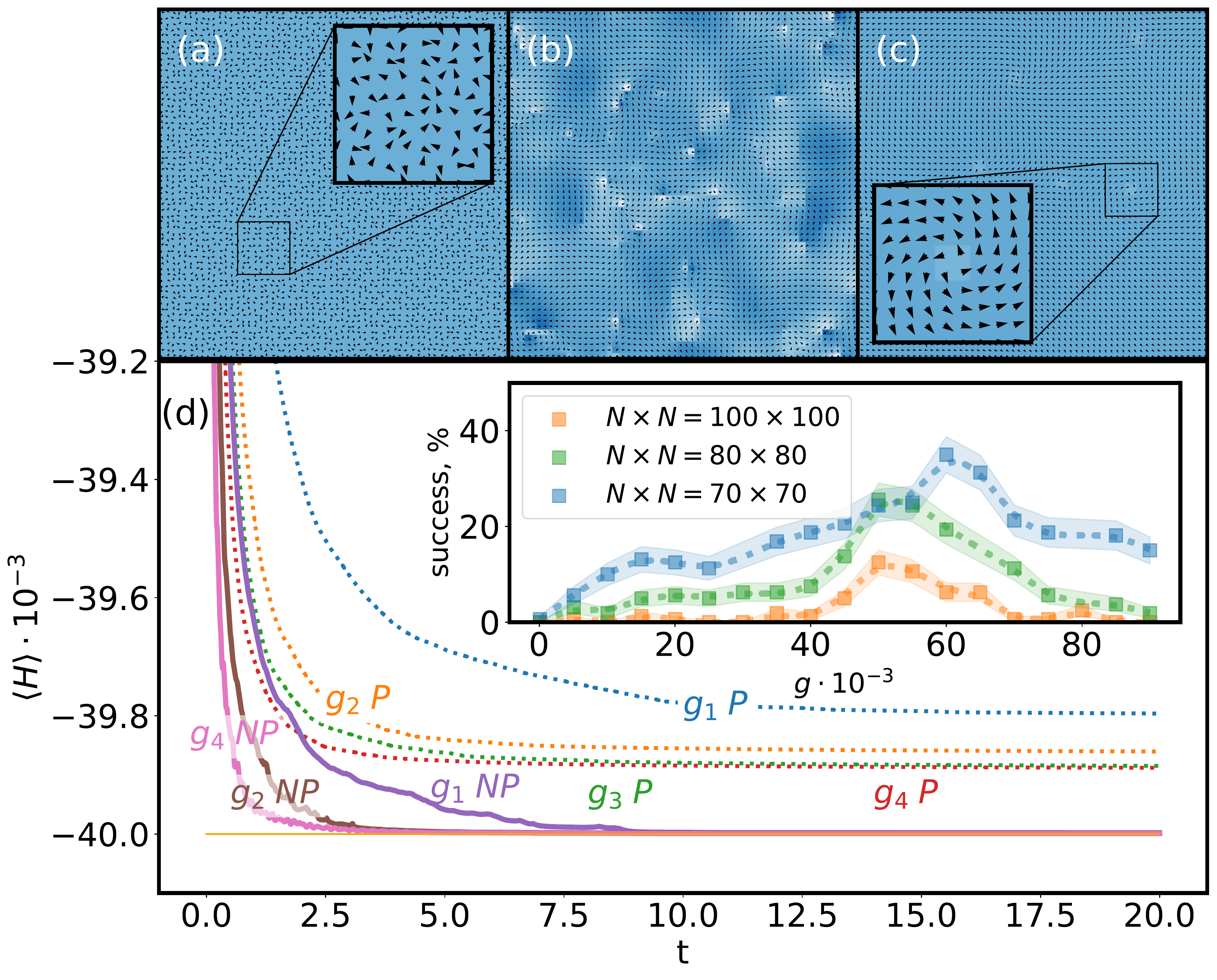}

        \caption{ Ferromagnetic couplings $J_{ij}=1$. (a - c) The time snapshots of the evolution of Eq.~(\ref{main}) on a 2D  square $100\times 100$ lattice  with NN planar interactions from (a) the randomly distributed initial state with $|\psi_{i}|=1$ and uniformly distributed $\phi_{i}$ (shown as black arrows)  to (c) the final steady-state with the persistence of isolated vortices. (d) Average Hamiltonian time evolution for 100 different initial conditions and  various strengths of nonlinearity  $g$. Label $P$ [$NP$] implies planar NN [non-planar] couplings  discussed in the main text. Here $g_1 = 10^2$, $g_2 = 10^3$, $g_3 = 2 \cdot 10^3$, $g_4 = 6 \cdot 10^3$.  (inset of d) The success probability of  finding the ground  (vortex-free) state during the evolution  for $K=100$ as a function of $g$. Shading represents the variability of the result for different runs.  }
        \label{fig:AverageH}
    \end{figure}

   {\it Low-dimensional topologies of coupling matrix.} The formation of domains and topological structures on the route to GS is well understood for  the classical lattice models with $J_{ij}=J>0$  in the context of  BEC formation (quench dynamics) from a strongly degenerate gas of weakly interacting bosons \cite{berloff2002scenario, weiler2008spontaneous, sun2017bose}. 
    In toroidal  1D geometry, a global and persistent phase winding can spontaneously form \cite{das2012winding}. In 2D, the establishment of the long-range order and  a genuine condensate in GS is replaced by the persistence of the isolated vortices according to the  Berezinskii–Kosterlitz–Thouless (BKT) transition scenario  \cite{berezinskii1972destruction,kosterlitz1973ordering}.  We conclude  that the existence of fully or partially disjoined toroidal 1D or planar subsets in the adjacency matrix ${\bf J}$ prevents the system from achieving its GS. In toroidal 1D configurations, the global phase windings  characterize the excited states of the system, while topological protection prohibits  further transition to the GS.  Strongly converging vortex-antivortex pairs (VAPs) in 2D can eventually annihilate, bringing the system to the vortex-free GS. However, the larger is  effective separation of vortices, the less likely they interact, the more likely the 2D subsystem remains in the excited state. The detailed analysis of these two scenarios is presented in the Supplementary Information (SI), where we also studied their strong dependence on nonlinearity $g$ and the rate of dissipation $K$. Both of these parameters can vary from system to system and often can be controlled within one system (e.g. using Feshbach resonance in ultracold atomic BECs to change $g$ \cite{bauer2009control} or the cooling rate to control $K$).) The effective vortex core size depends on $g$ as $\xi \sim (J/g)^{1/2}$ which sets a competition between two opposing effects. Stronger nonlinearity leads to stronger interactions and, therefore, more efficient vortex annihilation. On the other hand, sufficiently large $g$  reduces the scale at which isolated vortices can still interact with each other, which leads to  {\it freezing}  of plane vortices; see Fig.~\ref{fig:AverageH}(c,d). Moreover, for each size of the 2D subsystem, there is an optimal $g$ that maximizes the success probability for finding GS; see Fig.~\ref{fig:AverageH}(the inset of d).
   
\begin{figure}[t]
        \centering
        \includegraphics[width=\linewidth]{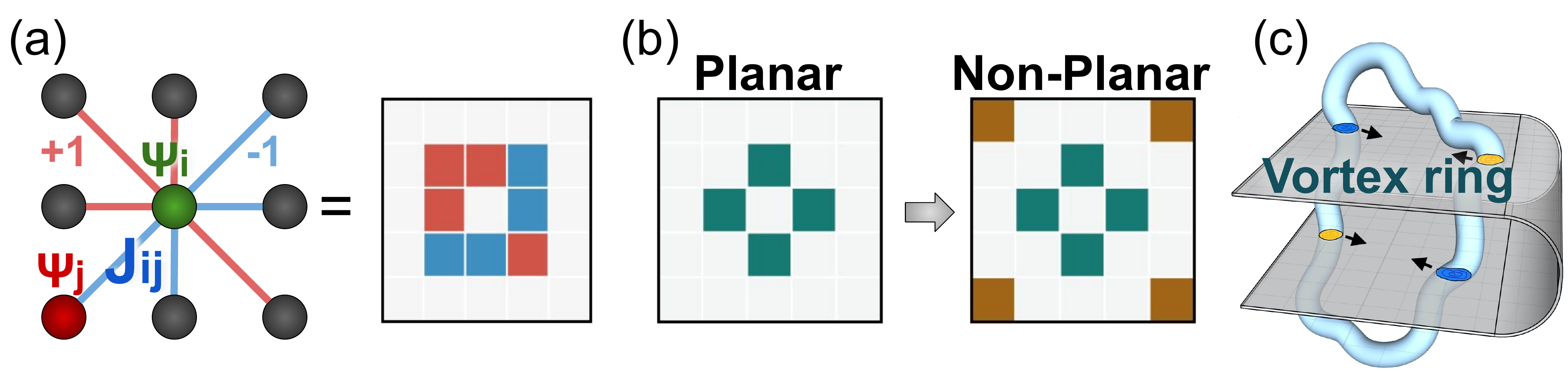}\hfil
        
        \caption{ 
        (a) Representation of interactions of individual node $\psi_i$ according to  ${\bf S}$. Red [blue] edges/cells represent ferromagnetic [anti-ferromagnetic] interactions with the central cell.
        (b) An illustration of  one non-planar coupling used in this paper (brown cells represent the additional non-planar  interactions).
        (c) Schematic representation of interacting VAPs on a quasi-3D lattice with non-planar couplings.}
        \label{fig:Sketch}
    \end{figure}

The distinctive feature of the quench dynamics of 3D systems is the formation of vortex filaments emerging from the local symmetry breaking \cite{berloff2002scenario}. The closed filaments -- vortex rings-- are the torus-shaped regions of phases rotating by $2\pi$ around a closed contours.  When the system evolves according to  Eq.~(\ref{main}), the vortex rings decay bringing the system to the true GS without any  topological defects persisting at the final state. This suggests that we can help the system achieve its GS by temporarily introducing non-planar (quasi-3D) couplings that prevent the appearance of the global winding and turn isolated VAPs into vortex rings to enable them to shrink annihilate. To describe this procedure, we consider
     the graph configuration of the interacting spins  $G = \big( U, V \big)$, as shown in Fig. \ref{fig:Sketch}. 
   The nodes of $G$ are characterised by  $\psi_i \in U$  while $(\psi_i, \psi_j) \in V$ represent the edges; see Fig.~\ref{fig:Sketch}(a). ${\bf J}$ becomes  the adjacency matrix of $G$. We define the coupling {\it scheme} $S^k = \big( U^k , V^k \big)$ of the individual node $\psi_k$ as the subgraph of G containing all nodes $\psi_j \in U^k$ adjacent to $\psi_k$ and all edges $(\psi_k, \psi_j) \in V^k$ between them; see Fig.~\ref{fig:Sketch}(a). We  say that graph $G$ and the corresponding ${\bf J}$ are generated by the scheme ${\bf S}$ if each individual node of $G$ can be described by ${\bf S}$.

    In 2D square geometry and NN interactions, $G$ is planar. To implement a quasi-3D geometry we introduce additional non-planar interactions shown in Fig.~\ref{fig:Sketch}(b). With these couplings, $G$ becomes non-planar and VAPs are replaced with vortex filaments,
    as schematically illustrated in Fig.~\ref{fig:Sketch}(c).
    Under the action of non-energy preserving dissipation in Eq.~(\ref{main}), the radius $R(t)$ of 3D vortex rings  obeys
   $R'(t) = - \frac{1}{2} a(K) \log(8 R(t))/R(t),$
    where $a(K)$ is the rate of the vortex ring dissipation \cite{Groszek_2018, berloff2007dissipative}.
    An average radius of the ring arising during the quench  with linear dimensionless domain size $L/\xi$ is  $R_0 = L/8\xi$. 
    To estimate the characteristic time for VAPs to pass through the ferromagnetic quasi-3D system, we place  VAP with separation $2R_0$   and numerically calculate $a(K)$ from Eq.~(\ref{main}) for 
     different values of $K$ and $g$ as 
    $a(K) =  R_0^2/{ \log(8 R_0) T},$ where $T$ is the dimensionless time it takes  to completely annihilate VAP. Here,  we used the result that the square of the vortex line length decays linearly with time \cite{berloff2007dissipative}. 
    We determined the  phenomenological form  of $a(K)$ as $a(K) = 1.35 K + 50.5$ (the fit of this formula to the numerical data is   shown in Fig.~S.5). The VAP dissipation rate $a(K)$ does not depend on $g$, which supports our assumption that VAP motion in quasi-3D geometry resembles the motion of the vortex rings in 3D.

    
    This suggests that to help the system to reach the ground state, we can introduce   additional non-planar couplings $B_{ij}$  (see Fig.~\ref{fig:Sketch}(b)) by replacing planar $J_{ij}$ with   $J_{ij} + \eta(t) B_{ij},$
    where 
    $\eta(t)$ is a monotonically decaying function of $t$, chosen here as the Fermi function $\eta(t) = (\exp[5(t - t_0 + 1)]+1)^{-1}.$
    For sufficiently long  attenuation times $t_0$ ($\sim T$), all VAPs annihilate and  bring the system to the GS with 100\% probability; see Fig.~\ref{fig:AverageH}(d) for the characteristic time evolution of $H$.

\begin{figure}[t]
         \centering
         \includegraphics[width=\linewidth]{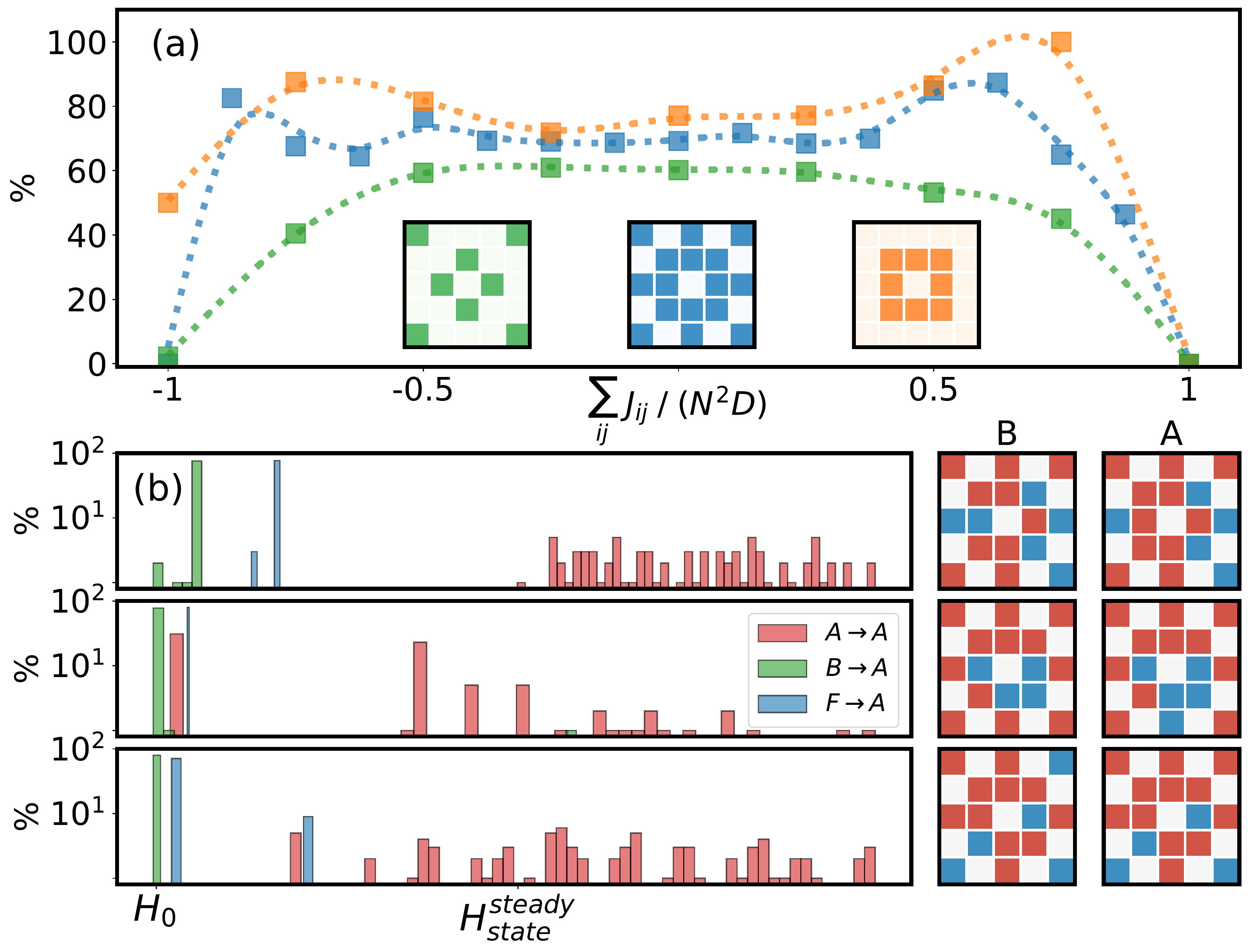}
         \caption{(a) Probability distribution for VAPs being trapped in different spin glass lattice models.  Mobility of VAPs is analysed by numerical integration of Eq.~(\ref{main}) with $K=100$, $g=500$  on $40\times 40$ doubly-periodic 2D square lattices using all possible $2^{16} + 2^8 + 2^8$ couplings ${\bf J}$  generated by schemes shown in the inset and using $50$  random VAP placements for each scheme.
         (b) Energy states found by three different annealing schemes starting from 80 random initial configurations and numerical integration of Eq.~{\ref{main}}.  "$A$" indicates the original glassy problem to be solved. $(X \rightarrow A)$ shows the distribution of the found energy states using the quenching process (if $X=A$), using  annealing from the ferromagnetic state (if $X=F$), and using the QVMA scheme (if $X=B$). Red [blue] cells in annealing schemes represent ferromagnetic [anti-ferromagnetic] interactions with the central cell.  }
         \label{fig:CouplingsDistr}
     \end{figure}

    We fulfilled the first task on the proposed route to find the GS of the general coupling matrix that consists of using quench dynamics while changing the system's topology to allow the topological defects to dissipate and annihilate while propagating through the system. However, for the general (sign and magnitude changing) form of adjacency matrix ${\bf J}$ 
     the potential energy landscape becomes rough so that topological structures are likely to become trapped in one of many local minima ({\it traps}), manifesting massive complexity of the problem of finding  GS for spin glasses. However, not every glassy system traps excitations. For these systems, the vortex phase imprinting  (or quench with vortices spontaneously generated)  brings the system to the GS, as we illustrate in the SI. So next, we develop an annealing scheme that allows excitations to escape the {\it traps}. We consider the local mobility of VAPs  in the neighborhood of each node. If VAPs are mobile, no changes to couplings are necessary during the annealing phase. If VAPs get trapped around some $\psi_i$, the task is to find the minimum set of edges connected to $\phi_i$  that allow VAPs to pass when changed.  To fully characterize all possible states, we focus on physically relevant couplings that are generated by schemes ${\bf S}$ shown in the inset to Fig.~\ref{fig:CouplingsDistr}(a) and  $J_{ij}=\pm J$. That gives us  $2^{16} + 2^{8} + 2^{8}$ possible coupling configurations.
     
     The coupling matrices ${\bf J}$ have been constructed by the following rule:
    \vspace{-0.3cm}
     \begin{align}
        &\hat{R}(Y) S^{(X,Y)}  \hat{R}(X) 
          =  \begin{bmatrix}
            0 &   & 1\\
              & \udots &  \\
            1 &   & 0
        \end{bmatrix}^Y
        \overbrace{
        \begin{gathered}
        \vspace{-0.1cm}
        \includegraphics[height=1.1cm]{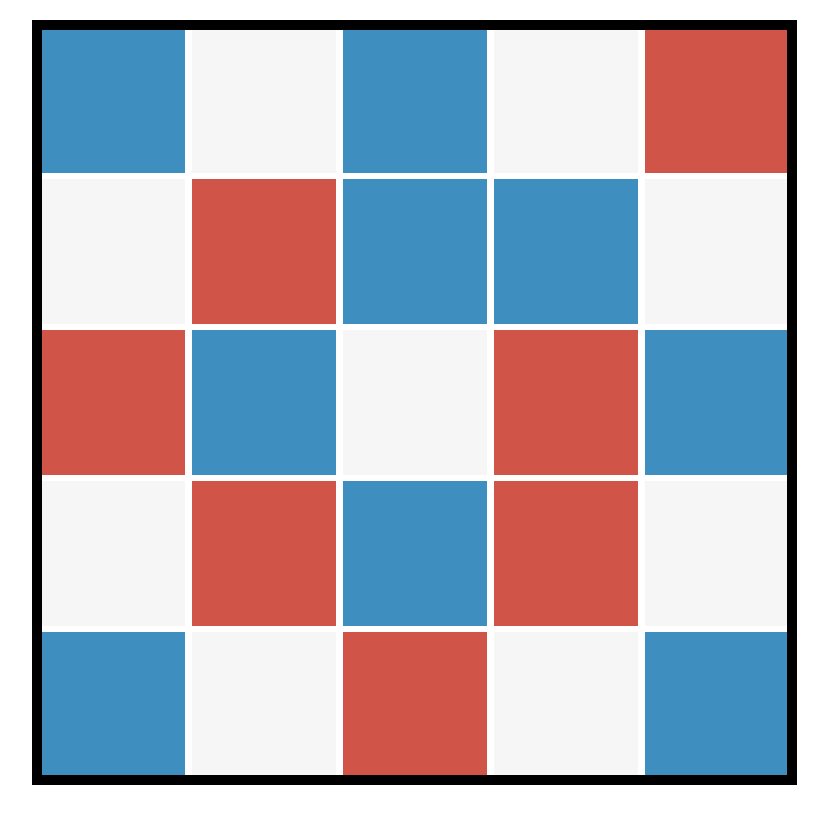}
        \end{gathered}
        }^{S}
        \;
        \begin{bmatrix}
            0 &   & 1\\
              & \udots &  \\
            1 &   & 0
        \end{bmatrix}^X, &
        \\
        &\hat{R}(P) = 
        \begin{bmatrix}
            0 & 0 & 1\\
            0 & \hat{I}_{3 \times 3} & 0 \\
            1 & 0 & 0
        \end{bmatrix}^{ F (P)} \quad F(P) = \frac{P - P \; \rm{mod} \; 2}{2}. &
        \nonumber 
        \end{align}
        \label{eq:GenRule} 
    Here $S^{(X,Y)}$ is a  coupling  scheme connecting  $\psi{(X,Y)} \in G$  with its neighbours  with coordinates denoted by $X,Y \in \mathbb{N}$ (Eq.~(5) uses one example of such $S$). 
    
    Firstly, we need to determine the mobility of VAPs for different schemes. To do this, we prepare each $40\time 40$ doubly-periodic lattice in the steady-state with some low energy $H_0$ (this can be done by integrating Eq.~(\ref{GP}) in imaginary time  or using the gain-dissipative networks \cite{kalinin2018networks}). We inject a VAP at  random positions $(X_i,Y_i)$ by multiplying $\psi(X,Y)$ by $v(X-X_0,Y-Y_0)v^*(X-X_1,Y-Y_1)$ where $v(X,Y)=[X + i Y]/[X^2 + Y^2 + \xi]$ and $(X_1-X_0)^2+(Y_1-Y_0)^2=10^2$ \cite{berloff2004pade}. We numerically integrate Eq.~(\ref{main}) until the steady state is found with energy $H_i$.  If $H_i<H_0$, we redefine as $H_0=H_i$ for the rest of the calculations.  For each random VAP placement, VAPs either become trapped  ($H_i>H_0$ for some $i$ ) or not ($H_i\le H_0$ for all $i$). 
     Figure \ref{fig:CouplingsDistr}(a) depicts the probability of VAPs to be trapped somewhere on the lattice  as the function of   $\sum_{i,j} J_{ij}/N^2D$ remarkably illustrating high probability (around 60\%) of VAPs being trapped for any configuration away from either predominantly ferromagnetic or  antiferromagnetic couplings.

     Secondly, to demonstrate  our quantized vortex mediated annealing (QVMA) approach, we selected the set $A$ of  {\it worst} coupling schemes denoted by $A_i$ with $D=16$ that trap VAPs with 100\% probability.  We denote by $B=\{ \{ B_j \}_i \}$ the sets of coupling schemes where VAPs are free to move while  each $B_j \in \{ B_j \}_i$ differs from $A_i \in A$ by a single coupling strength. In total there were $|A|=3545$ and $|B|=5418$ coupling schemes with this property.  We study the quench dynamics according to  Eq.~(\ref{main}) starting from  $\{ \{ B_j \}_i \}$ and slowly (on the timescale of the vortex annihilation $T$) transform couplings from one such set  (when several sets satisfy this property we will take the one with the minimum energy configuration reached at the end of the annealing). Such annealing transformation  is realised by $J_{ij} =J^{B/F}_{ij}$ if  $t < T_i$, $J_{ij} =( 1 - (t - T_i)/\Delta T) J^{B / F}_{ij} + (t - T_i)J^{A}_{ij}/\Delta T $ if $T_i \leq t\leq  T_f$, and $J_{ij} =J^{A}_{ij}$ if $t > T_f.$
    Here, $\Delta T=T_f - T_i$, $T_i$ and $T_f$ represent the beginning and the end moment of annealing: $T_i \approx T$ is the time required to reach a steady-state, and $T_f \approx 15T$.  $J_{ij}^{A[B/F]}$ are the couplings given by scheme $A_i$ [$B_i$ or all ferromagnetic couplings.] For such set $A$, we calculated the fraction of matrices for which  the lowest energy state within 0.5\% of the corresponding spectral width was found using 80 random initial states per each matrix and three methods: (1) QVMA ($B\rightarrow A$); (2) quench dynamics without annealing ($A\rightarrow A$) and (3) from the  ferromagnetic state with annealing ($F\rightarrow A$). QVMA ($B\rightarrow A$) achieved minimization of 76\% of all $3545$ matrices, compared with 13\% for  $F\rightarrow A$ and 44\% for $A\rightarrow A$. 
    The success probabilities of finding various energy states from 80 different initial configurations are depicted in Fig.~\ref{fig:CouplingsDistr}(b)  for three different schemes $A_i$ demonstrating a striking improvement of QVMA over other methods. 
    
    {\it  In summary}, we argued that two main characteristics of the coupling graph of the universal complex matter field lattice systems prevent it from obtaining GS during the dissipative quench dynamics: (i) the presence of low (one- or two-) dimensional subgraphs that leads to either global winding or isolated noninteracting and, therefore, non-dissipative topological defects and (ii) the existence of trapping regions for mobile excitations. We have shown that both characteristics can be efficiently annealed from the modified coupling graphs with favourable characteristics bringing the original system to the ground state with high probability. Such modified coupling graphs satisfy the following properties: they are (a)  non-planar (to facilitate quasi-3D vortex evolution); (b) do not trap vortex excitations; (c) have a minimum difference in couplings with the original graph. This is  the first time the engineered mobility of quantized vortices was used as a mediator in the system's transition to the GS. Our work suggests that bringing the graph topology, quench dynamics, and topological defects together may open new routes in global physically enhanced optimization.

\section{Supplementary Information: Quantized Vortices Mediated Annealing }

     \subsection{Role of nonlinearity and dissipation in one- and two-dimensional graph topologies}
     
\begin{figure}[!h]
        \centering
        \includegraphics[width=\linewidth]{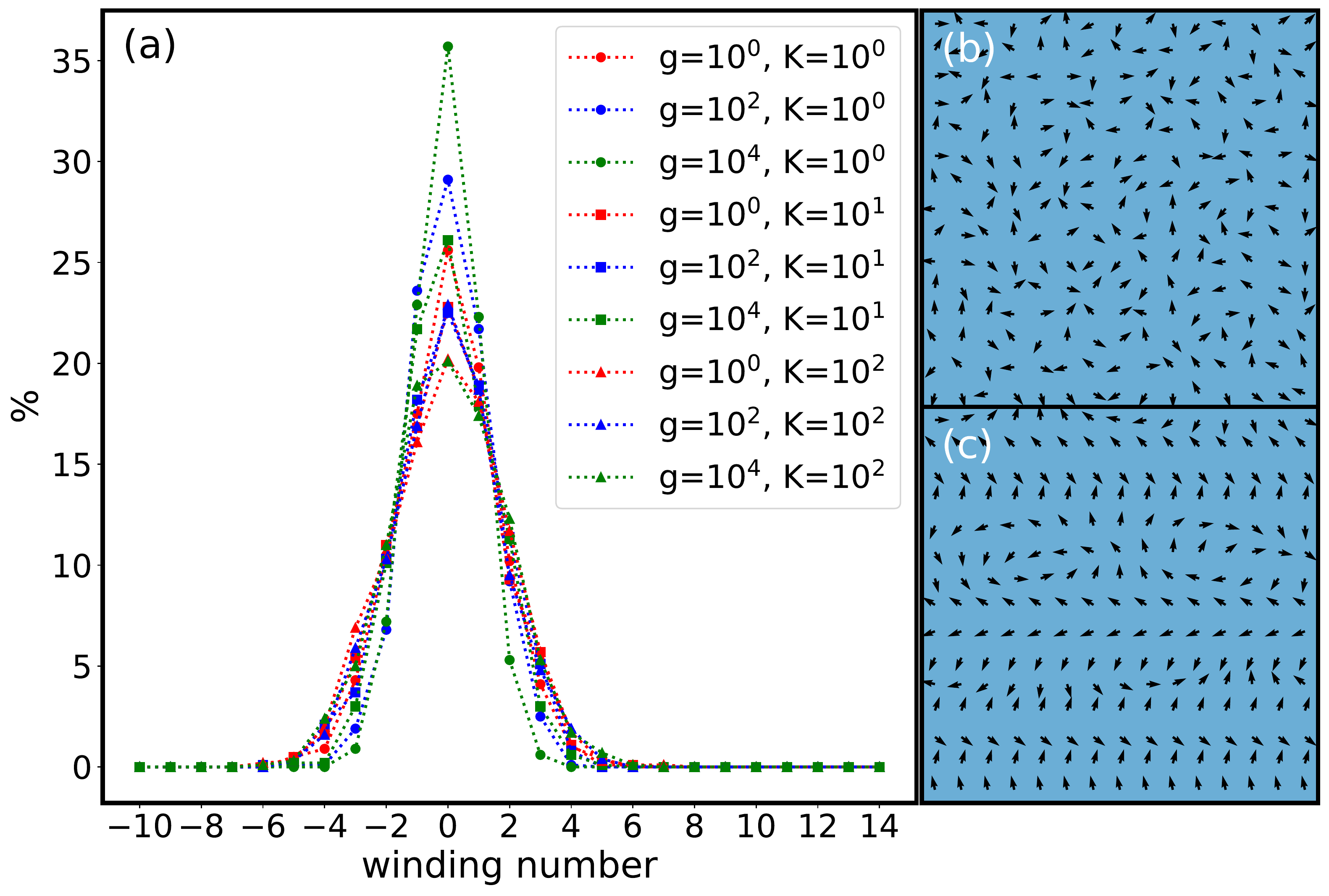}\hfil
        \caption{(a) Probability distribution of Winding Numbers obtained by numerical integration of Eq.~(M.4) (of the main text) on 1D lattice of size $N=100$ with NN interactions and various $g$ and $K$ parameters with the set of $1000$ randomly distributed initial states ($|\psi_i|=1$ and random phases $\phi_i \in [0,2\pi]$). (b) The phase representations of 10 horizontally oriented initial states. (c) The phase representations  of different  final states (depicted are only 10 of 1000). }
        \label{fig:1DWindings}
    \end{figure}

    Here we consider 1D and 2D periodic lattices of size $N=100$ and $N \times N = 100 \times 100$ respectively,  with purely ferromagnetic nearest neighbour interactions.
    We define the winding number $W$ of phase configuration as the total number of times that phases rotate by $2 \pi$ in positive (counter-clock) or negative (clockwise) directions. We numerically integrate Eq.~(M.4) of the main text starting with $|\psi_i|=1$ and randomly distributed phases of $\psi_i$. 
    The quenching dynamics in the one-dimensional periodic lattices can result in the ground state with zero phase differences between the neighboring sites so that $W = 0$ or in the formation of the global phase winding  $W \ne 0$ as shown in Fig.~\ref{fig:1DWindings}(c).
    Phase configurations with $W \ne 0$ characterize the excited states of the system, and topological protection prohibits their further transition to the ground state.
    We studied the numerical evolution of one thousand 1D spin-lattices (each of size $N=100$) with various $g$ and $K$ system parameters. Figure \ref{fig:1DWindings}(a) shows that the formation of the excited states with irreducible nonzero $W$ value strongly depends on nonlinearity of the self-interactions $g$ and the dissipation level $K$ of the system.
    Attainment of the final steady-state is estimated using the order parameter
    \begin{equation}
        O(t) = \bigg|\frac{1}{N} \sum_{i=1}^N \psi_i(t) \bigg|^2.
        \label{eq:order}
    \end{equation}
    The final steady-state is characterized by the stabilization of $O(t) \leq 1$ value in time. $O(t) = 1$ corresponds to the ground state.
    As shown  in Fig. \ref{fig:1DWindings} (a), the formation of stable excited states with $W \ne 0$ decreases with the increase of $g$ in 1D systems.
    The local symmetry breaking occurring in 2D lattices leads to the formation of plane vortices from domains with independent choices of broken symmetry.

    Strongly converging vortex-antivortex pairs can eventually annihilate, bringing the system to the vortex-free ground state.
    The rate of the system transition to the ground state strongly depends on their separation distances which are characterized by the vortex core sizes $\xi$ that are set by the balance between the interactions between the sites that scales  as $J/\xi^2$) and the potential energy of self-interactions characterised by $g$. Therefore, $\xi \sim (J/g)^{1/2}$.

     There is  a competition between two opposing effects: stronger nonlinearity leads to stronger interactions and therefore more efficient vortex annihilation. On the other hand, sufficiently large $g$ values reduce the scale at which isolated vortices can still interact with each other. For sufficiently large $g$ values and fixed lattice size this results in the final state with the presence of "frozen" plane vortices, as shown in Fig.~M.1 (c) of the main text.
     
    We verify this by starting from the set of 100 randomly distributed initial conditions with $|\psi_{i}| \equiv 1$, and numerically evolving Eq.~(M.4) (from the main text) up to the stabilization of $H(t)$ dependence in time. Averaging $H(t)$ dynamics over all 100 initial conditions, we provide $\langle H (t) \rangle$ distribution shown in Fig.~M.1 (d). 
    Comparing the final state Hamiltonian with the expected value of the ground state $H_{GS} = -\sum_{i j} J_{ij}$, we conclude whether the "frozen" vortices remain in the system by the end of the quenching process or not.
    Figure M.1 (d inset) depicts the obtained success probability distribution for achieving  the vortex-free ground state for different magnitudes of nonlinear self-interactions $g$ in the $100 \times 100$ system.
     
     Note that the periodic boundary conditions we used endow the lattice with a toroidal topological structure that has zero Euler characteristic $\chi$. According to the Poincaré–Hopf theorem, the sum of the singularity indices should satisfy the expression $\chi = \sum I_i = 0 $, where $I_i = {\rm sign} W$. Therefore, local violation of the $U(1)$ symmetry leads to the formation of an equal number of vortices and anti-vortices, which makes the annihilation of all vortex-anti-vortex pairs fundamentally achievable.

    \subsection{Time evolution of energy during the quench process in glassy systems with and without annealing}
    
\begin{figure}[!h]
         \centering
         \includegraphics[width=0.9\linewidth]{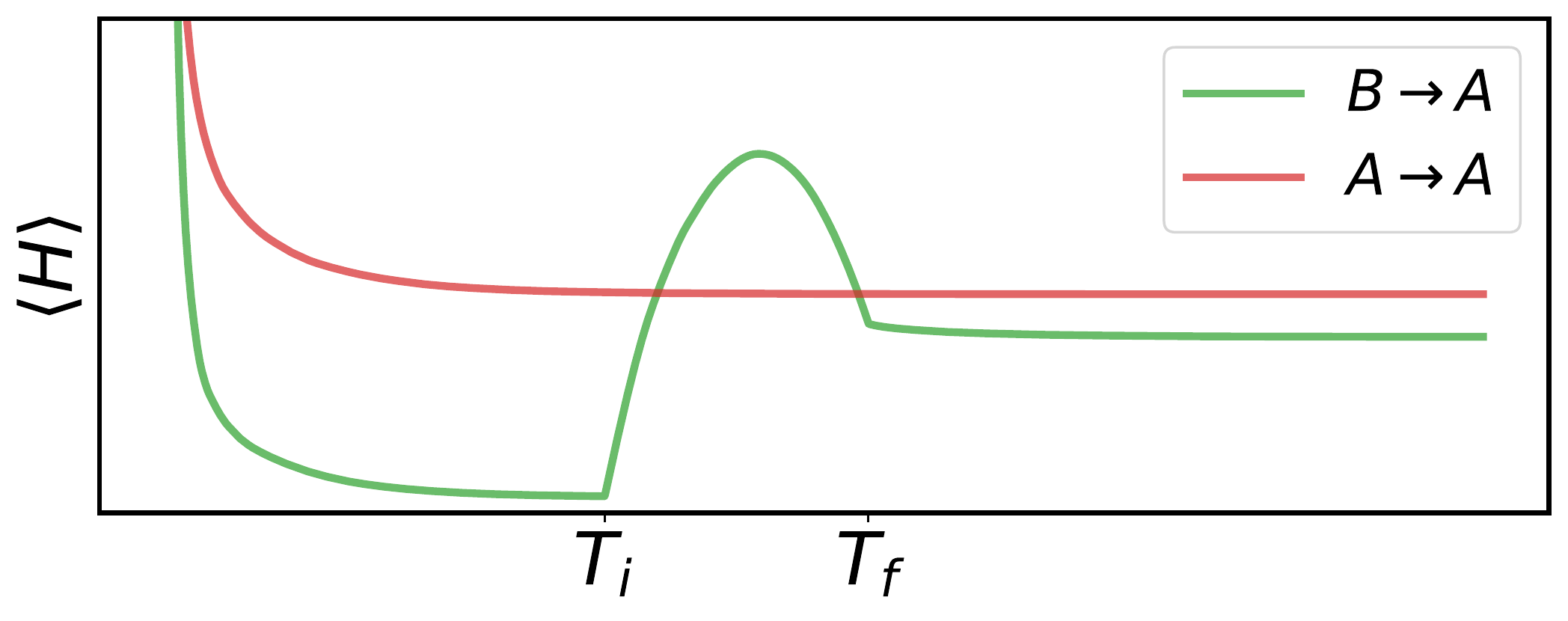}
         \caption{Schematics representation of the quench dynamics of the average energy of the system. 
        Time evolution of the system marked as $A\rightarrow A$ does not involve annealing procedure. The other one (marked as $B \rightarrow A $) starts with a coupling scheme $B$ that differs from $A$ by one coupling strength while admitting  free propagation of localised excitations through the system.  The adiabatic transformation of $B$ into $A$  takes place between $t=T_i$ and $t=T_f$ as discussed in the main text. Here $T_i \approx T$ denotes the time required for the system to pass to the steady state and $T_f \approx 15 \cdot T$.}
         \label{fig:Good2BadH}
     \end{figure}

    The  energy time dynamics   with and without annealing shows different stages  of the transformation. When spin-glass system undergoes the quench process, the energy monotonically decreases in time and achieves the excited state as the localised excited states dissipate and get trapped by the potential energy landscape. When QVMA is implemented, the energy of the system may increase when the annealing phase starts, but as the trapped localised states are allowed to escape  the system achieves a lower state at the end of the annealing process. Figure \ref{fig:Good2BadH} schematically illustrates this process. 
     
    \subsection{Effect of vortex imprinting on transition to a lower energy state}
\begin{figure}[!h]
        \centering
        \includegraphics[width=\linewidth]{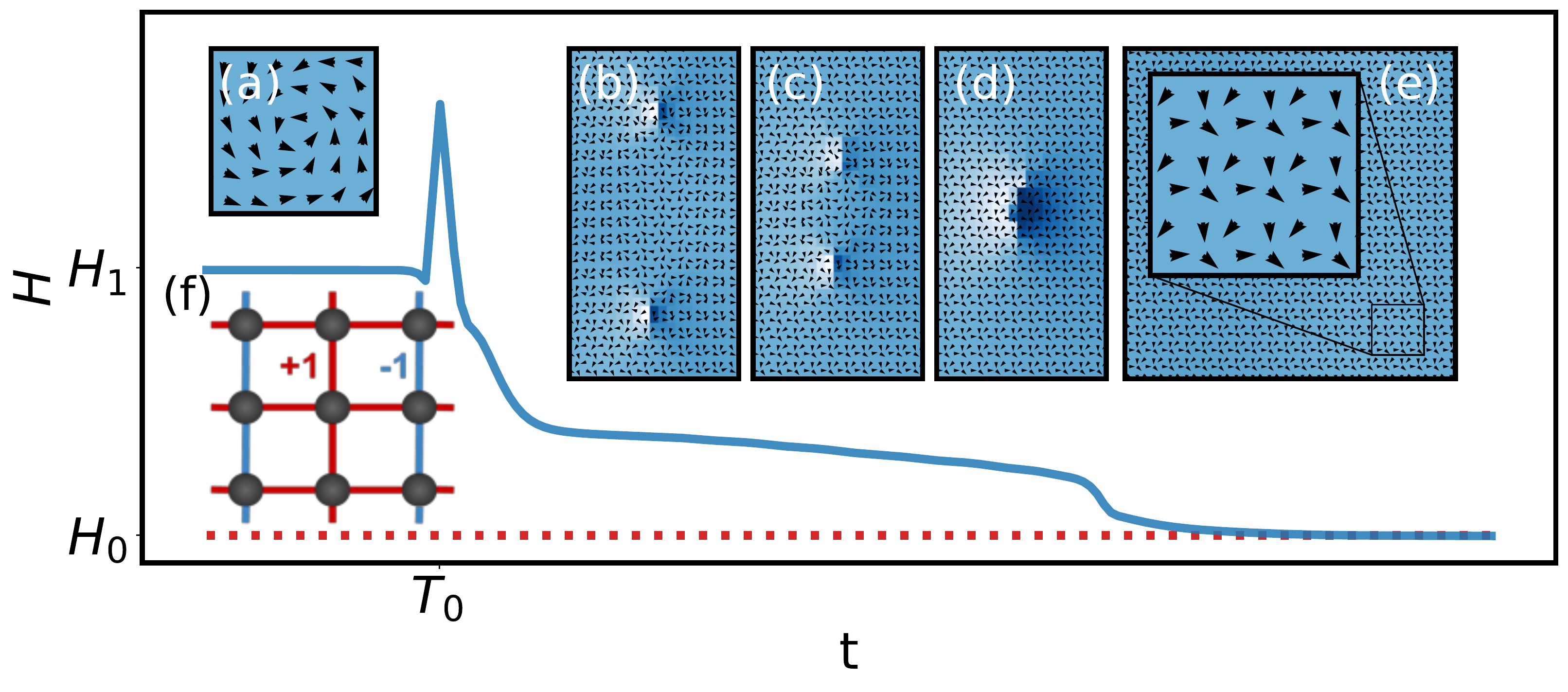}
        \caption{ Time-evolution of the energy (see Eq.~(M.1) (from the main text) of spin-glass of size $N\times N = 60 \times 60$ governed by Eq.~(M.4) (of the main text) with couplings given in the  inset (f) from uniformly distributed initial state $\psi_i \equiv 1$ and $K = 100, g = 500$.  The consecutive time snapshots of phases following the vortex pair injection at $t=T_0$ are given in (a - e). }
        \label{fig:AverageASH}
    \end{figure}

    Here we show that by introducing the vortex-antivortex pair we can bring a glassy system into a lower energy state.   Such imprinting can be realized experimentally, for example, using a laser beam \cite{white2014vortices}. Such imprinting is accompanied by an energy injection, which may be enough to overcome the local potential barrier and bring the system into a lower energy state. 
     To demonstrate this, we numerically integrate Eq.~(M.4) from the main text with $K=100$, $g=500$,  and $J_{ij}$ as shown in the inset to Fig.~\ref{fig:AverageASH}(f)  starting with $\psi_i(t=0)=1$. During the time evolution the system settles into an excited steady state with energy $H_1$. At time $T_0$ we introduce vortex-antivortex pair as discussed in the main text. This leads to a spike in the energy followed by a slow relaxation to the true ground state with energy $H_0$ as Fig.~\ref{fig:AverageASH} illustrates. 
    
     \subsection{Vortex trapping probability for other coupling rules}
\begin{figure}[!h]
         \centering
         \includegraphics[width=1\linewidth]{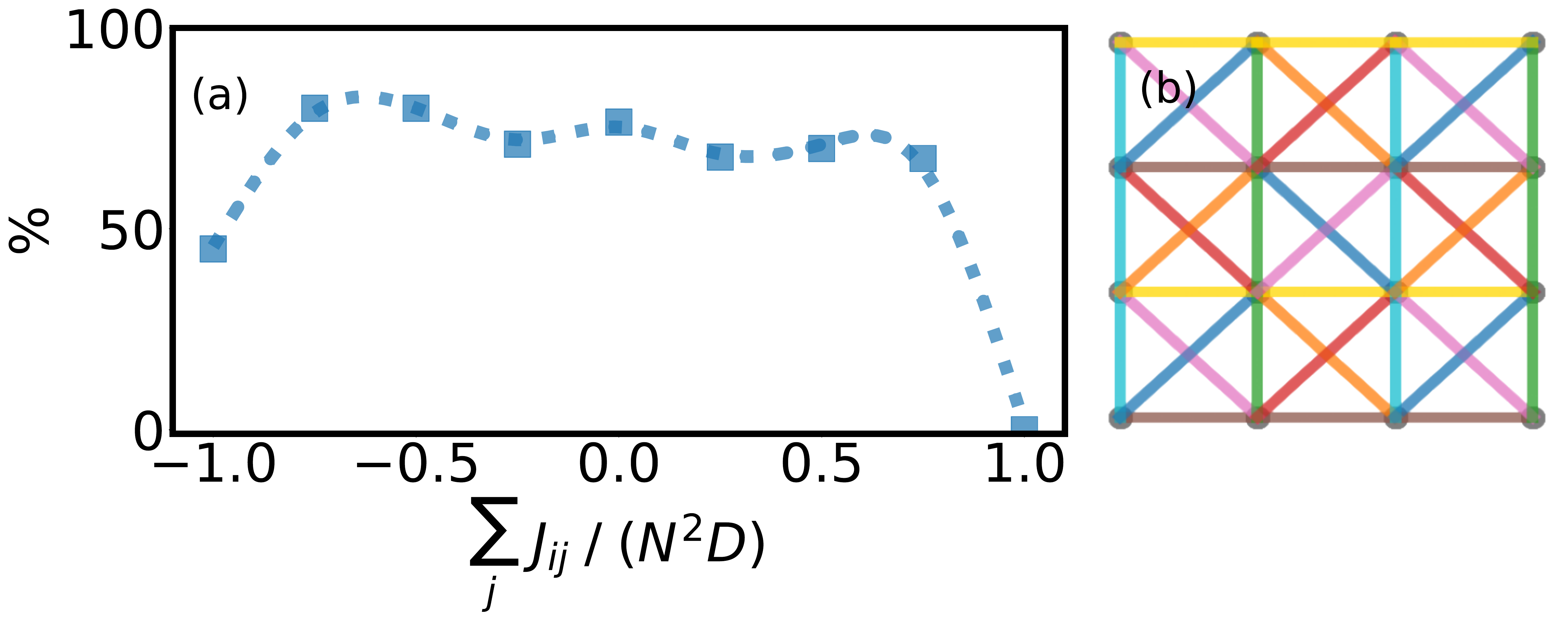}
         \caption{(a) The probabilities of vortex trapping in spin-glasses of size $N\times N = 40\times 40$ obtained by numerical numerical evolution of Eq.~(M.4) of the main text with the structure of interactions shown in (b) and $K=100, g=500$. Here, $D=8$ is the degree of each node, different colors correspond to different ferromagnetic or anti-ferromagnetic couplings, and edges with the same color have the same coupling strength. Probabilities has been calculated via $50$ vortex-antivortex pairs (VAPs) imprinting for each of $2^8$ possible $J_{ij}$ structures. Detailed description of how this distribution was obtained is given in the last part of the main text.}
         \label{fig:CouplingsDistrOld}
     \end{figure}

   Clearly,  Fig.~M.3 of the main text does not exhaust all possible types of couplings.  The rule we have chosen for constructing matrix $J_{ij}$, described by Eq.~(M.5) of the main text, is only one of many possible rules. However, the general characteristics of the vortex trapping probability we discussed in the main text are quite robust: the trapping probability saturates at about 60\%-70\% away from the couplings that are predominantly ferro- or antiferromagnetic as Fig.~\ref{fig:CouplingsDistrOld}(a) illustrates. The coupling strengths  represented in Fig.~\ref{fig:CouplingsDistrOld}(b) give another rule for constructing matrix $J_{ij}$. Here, the edges of same color have the same ferromagnetic or anti-ferromagnetic coupling strengths. For each of $2^{8}$  possible choices of $J_{ij}\in \{-1,1\}$ satisfying described rule, the system evolution from 50 different initial conditions  with a randomly located single vortex-antivortex pairs separated by distance $20$ on $40 \times 40$ lattice sites was investigated by the numerical integration of Eq.~(M.4) with $K=100$ and $g=500$.  
    
    \subsection{Evaluation of the vortex dissipation }
    \begin{figure}[!h]
         \centering
         \includegraphics[width=1\linewidth]{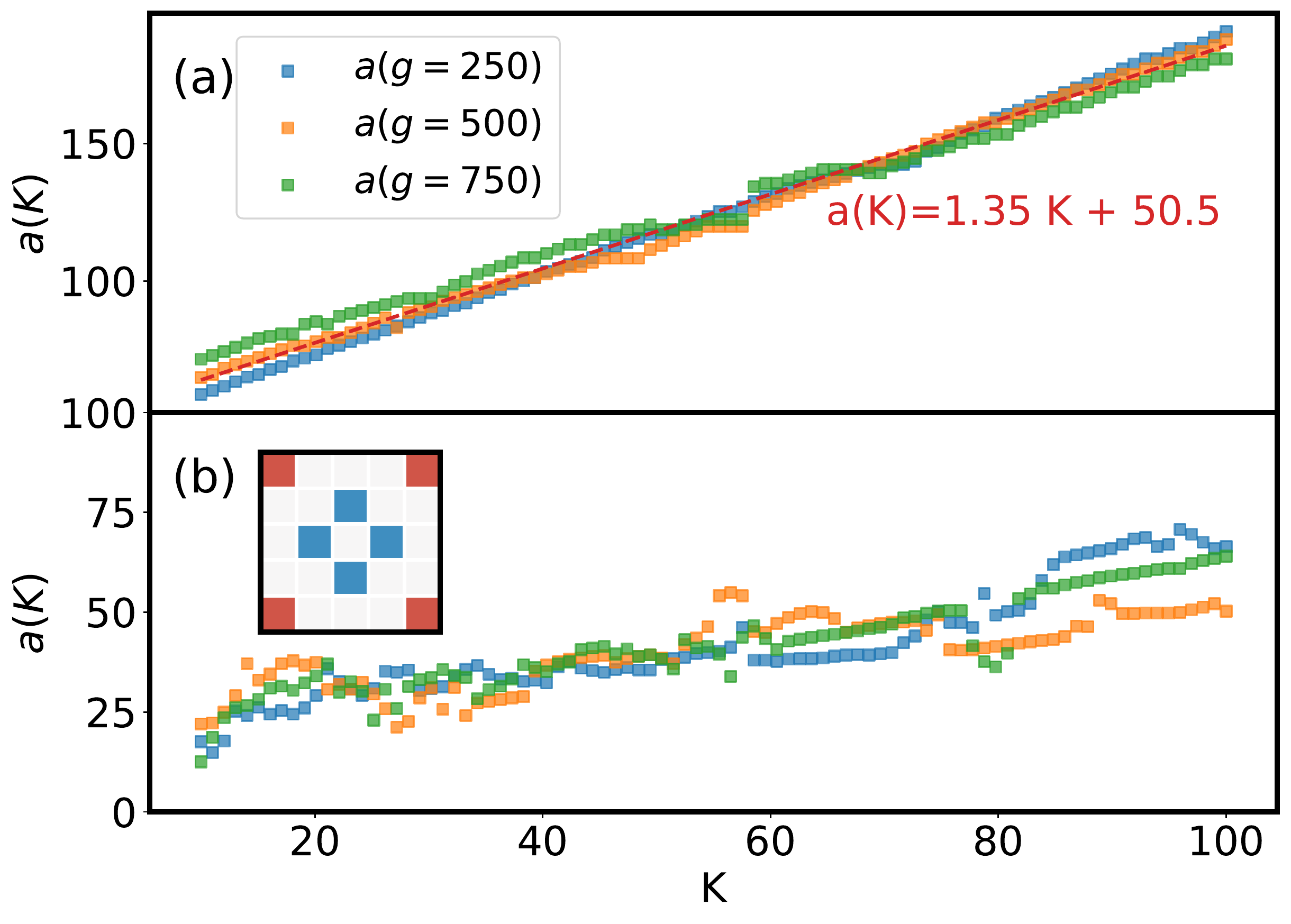}
         \caption{Vortex-antivortex pairs characteristic dissipation rate $a(K)$ calculated for different values of $g$ on the (a) ferromagnetic quasi-3D lattice (generated by scheme represented in Fig.~M.2(b) of the main text) and (b) glassy quasi-3D lattice (generated by scheme depicted in the (b inset)). The red dotted line represents the approximation fit of the $a(K)$ dependence cited in the main text which was obtained by numerical evaluation of the characteristic time $T$ required for injected VAPs governed by Eq.~(M.4) to converge and annihilate.}
         \label{fig:VortexShrinking}
     \end{figure}
     
     To quantify the dynamics of vortices on spin glasses, we evaluate the VAPs decay rate dependence $a(K)$ for glassy spin systems using the procedure presented  in the main text for ferromagnetic couplings; see Fig. \ref{fig:VortexShrinking}(a). The scheme with additional long-range non-planar interactions represented in Fig. \ref{fig:VortexShrinking}(b inset) endow the lattice with quasi-3D geometry. The lattice generated by this scheme does not contain any trapping regions for vortices. Therefore, we expect that nothing will prevent the vortices from passing through the lattice, and the system has to reach its global minimum eventually.
     Using the expression $a(K) =  R_0^2/{ \log(8 R_0) T}$ derived in the main text, we estimate the  characteristic time for the vortices to pass through the glassy system placing VAPs at a distance of $2 R_0$ from each other. Here $R_0$ denotes the average radius of the vortex ring arising in the purely ferromagnetic 3D system. Evolving Eq.~(M.4) of the main text in time we calculate the time $T$ required for the vortices to annihilate. Finally, we calculate  $a(K)$  for several $g$ values; see  Fig.~\ref{fig:VortexShrinking}(b).
     We conclude that the vortex decay rate dependence on dissipation level remains close to linear  for glassy quasi-3D systems in the absence of traps. However, the evolution proceeds much slower in comparison with the case of purely ferromagnetic systems considered previously and shown in Fig.~\ref{fig:VortexShrinking}(a).
     
     \subsection{Effect of the distance between trapping regions on the vortex dynamics}
\begin{figure}[!h]
         \centering
         \includegraphics[width=0.9\linewidth]{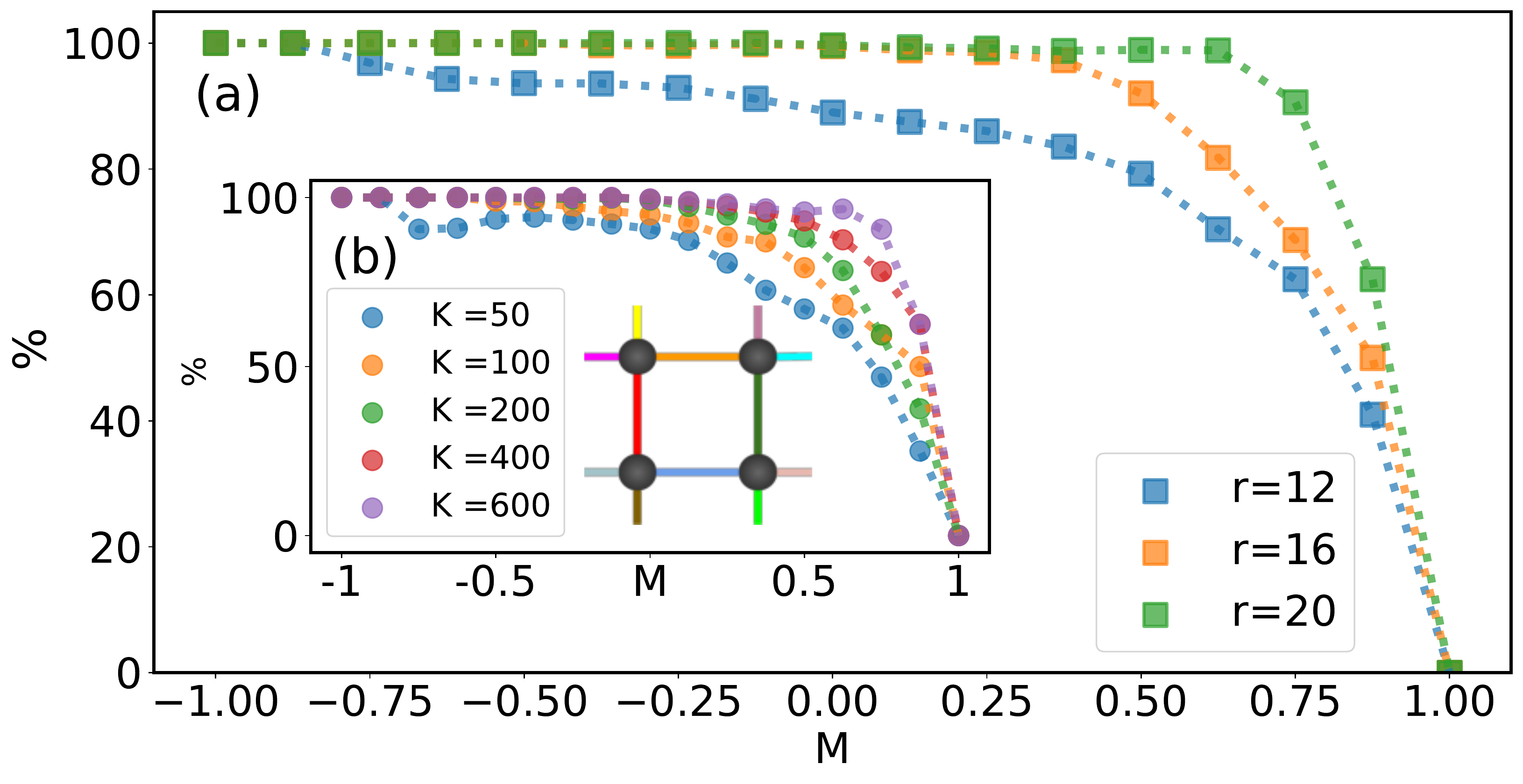}
         \caption{(a) The probabilities of VAPs  to get trapped on two plaquettes placed at a distance $r$ from each other, between  the VAPS and along the trajectory of their motion.  The plaquettes are the frustrated or non-frustrated square loops with the structure shown in (b inset). Colors depict twelve internal and external coupling strengths of the inserted plaquette, and each color can take either $+1$ or $-1$ value. Here $M$ is the total sum of the internal and external plaquette coupling strengths. For short separations $r$ the attraction of VAPs assists in the escape probability. (b) The dependence of the system's dissipation level $K$ on the trapping probabilities for plaquettes separated by $r=16$  and $g=500$.}
         \label{fig:Plackets}
     \end{figure}

     We studied the influence of the separation between the vortex traps on the capacity of vortices to avoid them. The simplest vortex trap can be realized by the insertion of the frustrated square loop into the lattice. We will call such frustrated or non-frustrated square loops represented in Fig.~\ref{fig:Plackets}(b inset) plaquettes. Twelve  internal and external coupling strengths of the inserted plaquette are depicted  in Fig.~\ref{fig:Plackets}(b inset), and coupling strength can take either $+1$ or $-1$ value. We denote by $M$ the total sum of the internal and external plaquette coupling strengths. Starting from the ground state of the purely ferromagnetic system with an imprinted VAP, we insert  two plaquettes of the same structure at a distance $r$ from each other between vortices. VAPs propagate through the system and  enter the plaquettes region. There they   can either be trapped or pass freely. Vortex trapping rate can be estimated by comparing the final state with the system's global minimum.  We evolve the Eq.~(M.4) with $K=200$ and $g=500$ numerically with all possible $2^{12}$ configurations of inserted plaquettes and different separations $r$ between them. The probabilities of vortices to be trapped by the inserted plaquettes are shown in Fig.~\ref{fig:Plackets}(a).  The smaller is the distance between the plaquettes the higher is the escape probability.

\bibliography{BibliographySimulators}{}

\begin{thebibliography}{41}%
\makeatletter
\providecommand \@ifxundefined [1]{%
 \@ifx{#1\undefined}
}%
\providecommand \@ifnum [1]{%
 \ifnum #1\expandafter \@firstoftwo
 \else \expandafter \@secondoftwo
 \fi
}%
\providecommand \@ifx [1]{%
 \ifx #1\expandafter \@firstoftwo
 \else \expandafter \@secondoftwo
 \fi
}%
\providecommand \natexlab [1]{#1}%
\providecommand \enquote  [1]{``#1''}%
\providecommand \bibnamefont  [1]{#1}%
\providecommand \bibfnamefont [1]{#1}%
\providecommand \citenamefont [1]{#1}%
\providecommand \href@noop [0]{\@secondoftwo}%
\providecommand \href [0]{\begingroup \@sanitize@url \@href}%
\providecommand \@href[1]{\@@startlink{#1}\@@href}%
\providecommand \@@href[1]{\endgroup#1\@@endlink}%
\providecommand \@sanitize@url [0]{\catcode `\\12\catcode `\$12\catcode
  `\&12\catcode `\#12\catcode `\^12\catcode `\_12\catcode `\%12\relax}%
\providecommand \@@startlink[1]{}%
\providecommand \@@endlink[0]{}%
\providecommand \url  [0]{\begingroup\@sanitize@url \@url }%
\providecommand \@url [1]{\endgroup\@href {#1}{\urlprefix }}%
\providecommand \urlprefix  [0]{URL }%
\providecommand \Eprint [0]{\href }%
\providecommand \doibase [0]{http://dx.doi.org/}%
\providecommand \selectlanguage [0]{\@gobble}%
\providecommand \bibinfo  [0]{\@secondoftwo}%
\providecommand \bibfield  [0]{\@secondoftwo}%
\providecommand \translation [1]{[#1]}%
\providecommand \BibitemOpen [0]{}%
\providecommand \bibitemStop [0]{}%
\providecommand \bibitemNoStop [0]{.\EOS\space}%
\providecommand \EOS [0]{\spacefactor3000\relax}%
\providecommand \BibitemShut  [1]{\csname bibitem#1\endcsname}%
\let\auto@bib@innerbib\@empty
\bibitem [{\citenamefont {Georgescu}\ \emph {et~al.}(2014)\citenamefont
  {Georgescu}, \citenamefont {Ashhab},\ and\ \citenamefont
  {Nori}}]{georgescu2014quantum}%
  \BibitemOpen
  \bibfield  {author} {\bibinfo {author} {\bibfnamefont {I.~M.}\ \bibnamefont
  {Georgescu}}, \bibinfo {author} {\bibfnamefont {S.}~\bibnamefont {Ashhab}}, \
  and\ \bibinfo {author} {\bibfnamefont {F.}~\bibnamefont {Nori}},\ }\href@noop
  {} {\bibfield  {journal} {\bibinfo  {journal} {Reviews of Modern Physics}\
  }\textbf {\bibinfo {volume} {86}},\ \bibinfo {pages} {153} (\bibinfo {year}
  {2014})}\BibitemShut {NoStop}%
\bibitem [{\citenamefont {Nielsen}\ and\ \citenamefont
  {Chuang}(2002)}]{nielsen2002quantum}%
  \BibitemOpen
  \bibfield  {author} {\bibinfo {author} {\bibfnamefont {M.~A.}\ \bibnamefont
  {Nielsen}}\ and\ \bibinfo {author} {\bibfnamefont {I.}~\bibnamefont
  {Chuang}},\ }\href@noop {} {\enquote {\bibinfo {title} {Quantum computation
  and quantum information},}\ } (\bibinfo {year} {2002})\BibitemShut {NoStop}%
\bibitem [{\citenamefont {Farhi}\ \emph {et~al.}(2001)\citenamefont {Farhi},
  \citenamefont {Goldstone}, \citenamefont {Gutmann}, \citenamefont {Lapan},
  \citenamefont {Lundgren},\ and\ \citenamefont {Preda}}]{farhi2001quantum}%
  \BibitemOpen
  \bibfield  {author} {\bibinfo {author} {\bibfnamefont {E.}~\bibnamefont
  {Farhi}}, \bibinfo {author} {\bibfnamefont {J.}~\bibnamefont {Goldstone}},
  \bibinfo {author} {\bibfnamefont {S.}~\bibnamefont {Gutmann}}, \bibinfo
  {author} {\bibfnamefont {J.}~\bibnamefont {Lapan}}, \bibinfo {author}
  {\bibfnamefont {A.}~\bibnamefont {Lundgren}}, \ and\ \bibinfo {author}
  {\bibfnamefont {D.}~\bibnamefont {Preda}},\ }\href@noop {} {\bibfield
  {journal} {\bibinfo  {journal} {Science}\ }\textbf {\bibinfo {volume}
  {292}},\ \bibinfo {pages} {472} (\bibinfo {year} {2001})}\BibitemShut
  {NoStop}%
\bibitem [{\citenamefont {Wilczek}(1982)}]{wilczek1982quantum}%
  \BibitemOpen
  \bibfield  {author} {\bibinfo {author} {\bibfnamefont {F.}~\bibnamefont
  {Wilczek}},\ }\href@noop {} {\bibfield  {journal} {\bibinfo  {journal}
  {Physical review letters}\ }\textbf {\bibinfo {volume} {49}},\ \bibinfo
  {pages} {957} (\bibinfo {year} {1982})}\BibitemShut {NoStop}%
\bibitem [{\citenamefont {Aharonov}\ \emph {et~al.}(1993)\citenamefont
  {Aharonov}, \citenamefont {Davidovich},\ and\ \citenamefont
  {Zagury}}]{aharonov1993quantum}%
  \BibitemOpen
  \bibfield  {author} {\bibinfo {author} {\bibfnamefont {Y.}~\bibnamefont
  {Aharonov}}, \bibinfo {author} {\bibfnamefont {L.}~\bibnamefont
  {Davidovich}}, \ and\ \bibinfo {author} {\bibfnamefont {N.}~\bibnamefont
  {Zagury}},\ }\href@noop {} {\bibfield  {journal} {\bibinfo  {journal}
  {Physical Review A}\ }\textbf {\bibinfo {volume} {48}},\ \bibinfo {pages}
  {1687} (\bibinfo {year} {1993})}\BibitemShut {NoStop}%
\bibitem [{\citenamefont {Bernstein}\ and\ \citenamefont
  {Vazirani}(1997)}]{bernstein1997quantum}%
  \BibitemOpen
  \bibfield  {author} {\bibinfo {author} {\bibfnamefont {E.}~\bibnamefont
  {Bernstein}}\ and\ \bibinfo {author} {\bibfnamefont {U.}~\bibnamefont
  {Vazirani}},\ }\href@noop {} {\bibfield  {journal} {\bibinfo  {journal} {SIAM
  Journal on computing}\ }\textbf {\bibinfo {volume} {26}},\ \bibinfo {pages}
  {1411} (\bibinfo {year} {1997})}\BibitemShut {NoStop}%
\bibitem [{\citenamefont {Jordan}\ \emph {et~al.}(2010)\citenamefont {Jordan},
  \citenamefont {Gosset},\ and\ \citenamefont {Love}}]{jordan2010quantum}%
  \BibitemOpen
  \bibfield  {author} {\bibinfo {author} {\bibfnamefont {S.~P.}\ \bibnamefont
  {Jordan}}, \bibinfo {author} {\bibfnamefont {D.}~\bibnamefont {Gosset}}, \
  and\ \bibinfo {author} {\bibfnamefont {P.~J.}\ \bibnamefont {Love}},\
  }\href@noop {} {\bibfield  {journal} {\bibinfo  {journal} {Physical Review
  A}\ }\textbf {\bibinfo {volume} {81}},\ \bibinfo {pages} {032331} (\bibinfo
  {year} {2010})}\BibitemShut {NoStop}%
\bibitem [{\citenamefont {Moussa}\ \emph {et~al.}(2010)\citenamefont {Moussa},
  \citenamefont {Ryan}, \citenamefont {Cory},\ and\ \citenamefont
  {Laflamme}}]{moussa2010testing}%
  \BibitemOpen
  \bibfield  {author} {\bibinfo {author} {\bibfnamefont {O.}~\bibnamefont
  {Moussa}}, \bibinfo {author} {\bibfnamefont {C.~A.}\ \bibnamefont {Ryan}},
  \bibinfo {author} {\bibfnamefont {D.~G.}\ \bibnamefont {Cory}}, \ and\
  \bibinfo {author} {\bibfnamefont {R.}~\bibnamefont {Laflamme}},\ }\href@noop
  {} {\bibfield  {journal} {\bibinfo  {journal} {Physical review letters}\
  }\textbf {\bibinfo {volume} {104}},\ \bibinfo {pages} {160501} (\bibinfo
  {year} {2010})}\BibitemShut {NoStop}%
\bibitem [{\citenamefont {Marandi}\ \emph {et~al.}(2014)\citenamefont
  {Marandi}, \citenamefont {Wang}, \citenamefont {Takata}, \citenamefont
  {Byer},\ and\ \citenamefont {Yamamoto}}]{marandi2014network}%
  \BibitemOpen
  \bibfield  {author} {\bibinfo {author} {\bibfnamefont {A.}~\bibnamefont
  {Marandi}}, \bibinfo {author} {\bibfnamefont {Z.}~\bibnamefont {Wang}},
  \bibinfo {author} {\bibfnamefont {K.}~\bibnamefont {Takata}}, \bibinfo
  {author} {\bibfnamefont {R.~L.}\ \bibnamefont {Byer}}, \ and\ \bibinfo
  {author} {\bibfnamefont {Y.}~\bibnamefont {Yamamoto}},\ }\href@noop {}
  {\bibfield  {journal} {\bibinfo  {journal} {Nature Photonics}\ }\textbf
  {\bibinfo {volume} {8}},\ \bibinfo {pages} {937} (\bibinfo {year}
  {2014})}\BibitemShut {NoStop}%
\bibitem [{\citenamefont {Nixon}\ \emph {et~al.}(2013)\citenamefont {Nixon},
  \citenamefont {Ronen}, \citenamefont {Friesem},\ and\ \citenamefont
  {Davidson}}]{nixon2013observing}%
  \BibitemOpen
  \bibfield  {author} {\bibinfo {author} {\bibfnamefont {M.}~\bibnamefont
  {Nixon}}, \bibinfo {author} {\bibfnamefont {E.}~\bibnamefont {Ronen}},
  \bibinfo {author} {\bibfnamefont {A.~A.}\ \bibnamefont {Friesem}}, \ and\
  \bibinfo {author} {\bibfnamefont {N.}~\bibnamefont {Davidson}},\ }\href@noop
  {} {\bibfield  {journal} {\bibinfo  {journal} {Physical review letters}\
  }\textbf {\bibinfo {volume} {110}},\ \bibinfo {pages} {184102} (\bibinfo
  {year} {2013})}\BibitemShut {NoStop}%
\bibitem [{\citenamefont {Berloff}\ \emph {et~al.}(2017)\citenamefont
  {Berloff}, \citenamefont {Silva}, \citenamefont {Kalinin}, \citenamefont
  {Askitopoulos}, \citenamefont {T{\"o}pfer}, \citenamefont {Cilibrizzi},
  \citenamefont {Langbein},\ and\ \citenamefont
  {Lagoudakis}}]{berloff2017realizing}%
  \BibitemOpen
  \bibfield  {author} {\bibinfo {author} {\bibfnamefont {N.~G.}\ \bibnamefont
  {Berloff}}, \bibinfo {author} {\bibfnamefont {M.}~\bibnamefont {Silva}},
  \bibinfo {author} {\bibfnamefont {K.}~\bibnamefont {Kalinin}}, \bibinfo
  {author} {\bibfnamefont {A.}~\bibnamefont {Askitopoulos}}, \bibinfo {author}
  {\bibfnamefont {J.~D.}\ \bibnamefont {T{\"o}pfer}}, \bibinfo {author}
  {\bibfnamefont {P.}~\bibnamefont {Cilibrizzi}}, \bibinfo {author}
  {\bibfnamefont {W.}~\bibnamefont {Langbein}}, \ and\ \bibinfo {author}
  {\bibfnamefont {P.~G.}\ \bibnamefont {Lagoudakis}},\ }\href@noop {}
  {\bibfield  {journal} {\bibinfo  {journal} {Nature materials}\ }\textbf
  {\bibinfo {volume} {16}},\ \bibinfo {pages} {1120} (\bibinfo {year}
  {2017})}\BibitemShut {NoStop}%
\bibitem [{\citenamefont {Cai}\ \emph {et~al.}(2019)\citenamefont {Cai},
  \citenamefont {Kumar}, \citenamefont {Van~Vaerenbergh}, \citenamefont {Liu},
  \citenamefont {Li}, \citenamefont {Yu}, \citenamefont {Xia}, \citenamefont
  {Yang}, \citenamefont {Beausoleil}, \citenamefont {Lu} \emph
  {et~al.}}]{memristors}%
  \BibitemOpen
  \bibfield  {author} {\bibinfo {author} {\bibfnamefont {F.}~\bibnamefont
  {Cai}}, \bibinfo {author} {\bibfnamefont {S.}~\bibnamefont {Kumar}}, \bibinfo
  {author} {\bibfnamefont {T.}~\bibnamefont {Van~Vaerenbergh}}, \bibinfo
  {author} {\bibfnamefont {R.}~\bibnamefont {Liu}}, \bibinfo {author}
  {\bibfnamefont {C.}~\bibnamefont {Li}}, \bibinfo {author} {\bibfnamefont
  {S.}~\bibnamefont {Yu}}, \bibinfo {author} {\bibfnamefont {Q.}~\bibnamefont
  {Xia}}, \bibinfo {author} {\bibfnamefont {J.~J.}\ \bibnamefont {Yang}},
  \bibinfo {author} {\bibfnamefont {R.}~\bibnamefont {Beausoleil}}, \bibinfo
  {author} {\bibfnamefont {W.}~\bibnamefont {Lu}},  \emph {et~al.},\
  }\href@noop {} {\bibfield  {journal} {\bibinfo  {journal} {arXiv preprint
  arXiv:1903.11194}\ } (\bibinfo {year} {2019})}\BibitemShut {NoStop}%
\bibitem [{\citenamefont {Kaminsky}\ \emph {et~al.}(2004)\citenamefont
  {Kaminsky}, \citenamefont {Lloyd},\ and\ \citenamefont
  {Orlando}}]{kaminsky2004scalable}%
  \BibitemOpen
  \bibfield  {author} {\bibinfo {author} {\bibfnamefont {W.~M.}\ \bibnamefont
  {Kaminsky}}, \bibinfo {author} {\bibfnamefont {S.}~\bibnamefont {Lloyd}}, \
  and\ \bibinfo {author} {\bibfnamefont {T.~P.}\ \bibnamefont {Orlando}},\
  }\href@noop {} {\bibfield  {journal} {\bibinfo  {journal} {arXiv preprint
  quant-ph/0403090}\ } (\bibinfo {year} {2004})}\BibitemShut {NoStop}%
\bibitem [{\citenamefont {Buluta}\ and\ \citenamefont
  {Nori}(2009)}]{buluta2009quantum}%
  \BibitemOpen
  \bibfield  {author} {\bibinfo {author} {\bibfnamefont {I.}~\bibnamefont
  {Buluta}}\ and\ \bibinfo {author} {\bibfnamefont {F.}~\bibnamefont {Nori}},\
  }\href@noop {} {\bibfield  {journal} {\bibinfo  {journal} {Science}\ }\textbf
  {\bibinfo {volume} {326}},\ \bibinfo {pages} {108} (\bibinfo {year}
  {2009})}\BibitemShut {NoStop}%
\bibitem [{\citenamefont {De~las Cuevas}\ and\ \citenamefont
  {Cubitt}(2016)}]{de2016simple}%
  \BibitemOpen
  \bibfield  {author} {\bibinfo {author} {\bibfnamefont {G.}~\bibnamefont
  {De~las Cuevas}}\ and\ \bibinfo {author} {\bibfnamefont {T.~S.}\ \bibnamefont
  {Cubitt}},\ }\href@noop {} {\bibfield  {journal} {\bibinfo  {journal}
  {Science}\ }\textbf {\bibinfo {volume} {351}},\ \bibinfo {pages} {1180}
  (\bibinfo {year} {2016})}\BibitemShut {NoStop}%
\bibitem [{\citenamefont {Svistunov}\ \emph {et~al.}(2015)\citenamefont
  {Svistunov}, \citenamefont {Babaev},\ and\ \citenamefont
  {Prokof'ev}}]{svistunov2015superfluid}%
  \BibitemOpen
  \bibfield  {author} {\bibinfo {author} {\bibfnamefont {B.~V.}\ \bibnamefont
  {Svistunov}}, \bibinfo {author} {\bibfnamefont {E.~S.}\ \bibnamefont
  {Babaev}}, \ and\ \bibinfo {author} {\bibfnamefont {N.~V.}\ \bibnamefont
  {Prokof'ev}},\ }\href@noop {} {\emph {\bibinfo {title} {Superfluid states of
  matter}}}\ (\bibinfo  {publisher} {Crc Press},\ \bibinfo {year}
  {2015})\BibitemShut {NoStop}%
\bibitem [{\citenamefont {Domb}(2000)}]{domb2000phase}%
  \BibitemOpen
  \bibfield  {author} {\bibinfo {author} {\bibfnamefont {C.}~\bibnamefont
  {Domb}},\ }\href@noop {} {\emph {\bibinfo {title} {Phase transitions and
  critical phenomena}}}\ (\bibinfo  {publisher} {Elsevier},\ \bibinfo {year}
  {2000})\BibitemShut {NoStop}%
\bibitem [{\citenamefont {Kevrekidis}\ \emph {et~al.}(2001)\citenamefont
  {Kevrekidis}, \citenamefont {Rasmussen},\ and\ \citenamefont
  {Bishop}}]{kevrekidis2001discrete}%
  \BibitemOpen
  \bibfield  {author} {\bibinfo {author} {\bibfnamefont {P.}~\bibnamefont
  {Kevrekidis}}, \bibinfo {author} {\bibfnamefont {K.}~\bibnamefont
  {Rasmussen}}, \ and\ \bibinfo {author} {\bibfnamefont {A.}~\bibnamefont
  {Bishop}},\ }\href@noop {} {\bibfield  {journal} {\bibinfo  {journal}
  {International Journal of Modern Physics B}\ }\textbf {\bibinfo {volume}
  {15}},\ \bibinfo {pages} {2833} (\bibinfo {year} {2001})}\BibitemShut
  {NoStop}%
\bibitem [{\citenamefont {Brazhnyi}\ and\ \citenamefont
  {Konotop}(2004)}]{brazhnyi2004theory}%
  \BibitemOpen
  \bibfield  {author} {\bibinfo {author} {\bibfnamefont {V.}~\bibnamefont
  {Brazhnyi}}\ and\ \bibinfo {author} {\bibfnamefont {V.}~\bibnamefont
  {Konotop}},\ }\href@noop {} {\bibfield  {journal} {\bibinfo  {journal}
  {Modern Physics Letters B}\ }\textbf {\bibinfo {volume} {18}},\ \bibinfo
  {pages} {627} (\bibinfo {year} {2004})}\BibitemShut {NoStop}%
\bibitem [{\citenamefont {Struck}\ \emph {et~al.}(2011)\citenamefont {Struck},
  \citenamefont {{\"O}lschl{\"a}ger}, \citenamefont {Le~Targat}, \citenamefont
  {Soltan-Panahi}, \citenamefont {Eckardt}, \citenamefont {Lewenstein},
  \citenamefont {Windpassinger},\ and\ \citenamefont
  {Sengstock}}]{struck2011quantum}%
  \BibitemOpen
  \bibfield  {author} {\bibinfo {author} {\bibfnamefont {J.}~\bibnamefont
  {Struck}}, \bibinfo {author} {\bibfnamefont {C.}~\bibnamefont
  {{\"O}lschl{\"a}ger}}, \bibinfo {author} {\bibfnamefont {R.}~\bibnamefont
  {Le~Targat}}, \bibinfo {author} {\bibfnamefont {P.}~\bibnamefont
  {Soltan-Panahi}}, \bibinfo {author} {\bibfnamefont {A.}~\bibnamefont
  {Eckardt}}, \bibinfo {author} {\bibfnamefont {M.}~\bibnamefont {Lewenstein}},
  \bibinfo {author} {\bibfnamefont {P.}~\bibnamefont {Windpassinger}}, \ and\
  \bibinfo {author} {\bibfnamefont {K.}~\bibnamefont {Sengstock}},\ }\href@noop
  {} {\bibfield  {journal} {\bibinfo  {journal} {Science}\ }\textbf {\bibinfo
  {volume} {333}},\ \bibinfo {pages} {996} (\bibinfo {year}
  {2011})}\BibitemShut {NoStop}%
\bibitem [{\citenamefont {Davis}\ \emph {et~al.}(2019)\citenamefont {Davis},
  \citenamefont {Bentsen}, \citenamefont {Homeier}, \citenamefont {Li},\ and\
  \citenamefont {Schleier-Smith}}]{davis2019photon}%
  \BibitemOpen
  \bibfield  {author} {\bibinfo {author} {\bibfnamefont {E.~J.}\ \bibnamefont
  {Davis}}, \bibinfo {author} {\bibfnamefont {G.}~\bibnamefont {Bentsen}},
  \bibinfo {author} {\bibfnamefont {L.}~\bibnamefont {Homeier}}, \bibinfo
  {author} {\bibfnamefont {T.}~\bibnamefont {Li}}, \ and\ \bibinfo {author}
  {\bibfnamefont {M.~H.}\ \bibnamefont {Schleier-Smith}},\ }\href@noop {}
  {\bibfield  {journal} {\bibinfo  {journal} {Physical review letters}\
  }\textbf {\bibinfo {volume} {122}},\ \bibinfo {pages} {010405} (\bibinfo
  {year} {2019})}\BibitemShut {NoStop}%
\bibitem [{\citenamefont {Gallina}\ and\ \citenamefont
  {Pastor}(2020)}]{gallina2020disorder}%
  \BibitemOpen
  \bibfield  {author} {\bibinfo {author} {\bibfnamefont {D.}~\bibnamefont
  {Gallina}}\ and\ \bibinfo {author} {\bibfnamefont {G.}~\bibnamefont
  {Pastor}},\ }\href@noop {} {\bibfield  {journal} {\bibinfo  {journal}
  {Physical Review X}\ }\textbf {\bibinfo {volume} {10}},\ \bibinfo {pages}
  {021068} (\bibinfo {year} {2020})}\BibitemShut {NoStop}%
\bibitem [{\citenamefont {Kadowaki}\ and\ \citenamefont
  {Nishimori}(1998)}]{kadowaki1998quantum}%
  \BibitemOpen
  \bibfield  {author} {\bibinfo {author} {\bibfnamefont {T.}~\bibnamefont
  {Kadowaki}}\ and\ \bibinfo {author} {\bibfnamefont {H.}~\bibnamefont
  {Nishimori}},\ }\href@noop {} {\bibfield  {journal} {\bibinfo  {journal}
  {Physical Review E}\ }\textbf {\bibinfo {volume} {58}},\ \bibinfo {pages}
  {5355} (\bibinfo {year} {1998})}\BibitemShut {NoStop}%
\bibitem [{\citenamefont {Hen}\ and\ \citenamefont
  {Sarandy}(2016)}]{hen2016driver}%
  \BibitemOpen
  \bibfield  {author} {\bibinfo {author} {\bibfnamefont {I.}~\bibnamefont
  {Hen}}\ and\ \bibinfo {author} {\bibfnamefont {M.~S.}\ \bibnamefont
  {Sarandy}},\ }\href@noop {} {\bibfield  {journal} {\bibinfo  {journal}
  {Physical Review A}\ }\textbf {\bibinfo {volume} {93}},\ \bibinfo {pages}
  {062312} (\bibinfo {year} {2016})}\BibitemShut {NoStop}%
\bibitem [{\citenamefont {Hen}\ and\ \citenamefont
  {Spedalieri}(2016)}]{hen2016quantum}%
  \BibitemOpen
  \bibfield  {author} {\bibinfo {author} {\bibfnamefont {I.}~\bibnamefont
  {Hen}}\ and\ \bibinfo {author} {\bibfnamefont {F.~M.}\ \bibnamefont
  {Spedalieri}},\ }\href@noop {} {\bibfield  {journal} {\bibinfo  {journal}
  {Physical Review Applied}\ }\textbf {\bibinfo {volume} {5}},\ \bibinfo
  {pages} {034007} (\bibinfo {year} {2016})}\BibitemShut {NoStop}%
\bibitem [{\citenamefont {Khalatnikov}(1965)}]{khalatnikov1965introduction}%
  \BibitemOpen
  \bibfield  {author} {\bibinfo {author} {\bibfnamefont {I.}~\bibnamefont
  {Khalatnikov}},\ }\href@noop {} {\bibfield  {journal} {\bibinfo  {journal}
  {Inc., New York}\ } (\bibinfo {year} {1965})}\BibitemShut {NoStop}%
\bibitem [{\citenamefont {De~Wijn}\ \emph {et~al.}(2013)\citenamefont
  {De~Wijn}, \citenamefont {Hess},\ and\ \citenamefont
  {Fine}}]{de2013lyapunov}%
  \BibitemOpen
  \bibfield  {author} {\bibinfo {author} {\bibfnamefont {A.}~\bibnamefont
  {De~Wijn}}, \bibinfo {author} {\bibfnamefont {B.}~\bibnamefont {Hess}}, \
  and\ \bibinfo {author} {\bibfnamefont {B.}~\bibnamefont {Fine}},\ }\href@noop
  {} {\bibfield  {journal} {\bibinfo  {journal} {Journal of Physics A:
  Mathematical and Theoretical}\ }\textbf {\bibinfo {volume} {46}},\ \bibinfo
  {pages} {254012} (\bibinfo {year} {2013})}\BibitemShut {NoStop}%
\bibitem [{\citenamefont {Tarkhov}(2020)}]{tarkhov2020}%
  \BibitemOpen
  \bibfield  {author} {\bibinfo {author} {\bibfnamefont {A.}~\bibnamefont
  {Tarkhov}},\ }\href@noop {} {\  (\bibinfo {year} {2020})}\BibitemShut
  {NoStop}%
\bibitem [{\citenamefont {Zong}\ \emph {et~al.}(2019)\citenamefont {Zong},
  \citenamefont {Kogar}, \citenamefont {Bie}, \citenamefont {Rohwer},
  \citenamefont {Lee}, \citenamefont {Baldini}, \citenamefont {Erge{\c{c}}en},
  \citenamefont {Yilmaz}, \citenamefont {Freelon}, \citenamefont {Sie} \emph
  {et~al.}}]{zong2019evidence}%
  \BibitemOpen
  \bibfield  {author} {\bibinfo {author} {\bibfnamefont {A.}~\bibnamefont
  {Zong}}, \bibinfo {author} {\bibfnamefont {A.}~\bibnamefont {Kogar}},
  \bibinfo {author} {\bibfnamefont {Y.-Q.}\ \bibnamefont {Bie}}, \bibinfo
  {author} {\bibfnamefont {T.}~\bibnamefont {Rohwer}}, \bibinfo {author}
  {\bibfnamefont {C.}~\bibnamefont {Lee}}, \bibinfo {author} {\bibfnamefont
  {E.}~\bibnamefont {Baldini}}, \bibinfo {author} {\bibfnamefont
  {E.}~\bibnamefont {Erge{\c{c}}en}}, \bibinfo {author} {\bibfnamefont {M.~B.}\
  \bibnamefont {Yilmaz}}, \bibinfo {author} {\bibfnamefont {B.}~\bibnamefont
  {Freelon}}, \bibinfo {author} {\bibfnamefont {E.~J.}\ \bibnamefont {Sie}},
  \emph {et~al.},\ }\href@noop {} {\bibfield  {journal} {\bibinfo  {journal}
  {Nature Physics}\ }\textbf {\bibinfo {volume} {15}},\ \bibinfo {pages} {27}
  (\bibinfo {year} {2019})}\BibitemShut {NoStop}%
\bibitem [{\citenamefont {Berloff}\ and\ \citenamefont
  {Svistunov}(2002)}]{berloff2002scenario}%
  \BibitemOpen
  \bibfield  {author} {\bibinfo {author} {\bibfnamefont {N.~G.}\ \bibnamefont
  {Berloff}}\ and\ \bibinfo {author} {\bibfnamefont {B.~V.}\ \bibnamefont
  {Svistunov}},\ }\href@noop {} {\bibfield  {journal} {\bibinfo  {journal}
  {Physical Review A}\ }\textbf {\bibinfo {volume} {66}},\ \bibinfo {pages}
  {013603} (\bibinfo {year} {2002})}\BibitemShut {NoStop}%
\bibitem [{\citenamefont {Weiler}\ \emph {et~al.}(2008)\citenamefont {Weiler},
  \citenamefont {Neely}, \citenamefont {Scherer}, \citenamefont {Bradley},
  \citenamefont {Davis},\ and\ \citenamefont
  {Anderson}}]{weiler2008spontaneous}%
  \BibitemOpen
  \bibfield  {author} {\bibinfo {author} {\bibfnamefont {C.~N.}\ \bibnamefont
  {Weiler}}, \bibinfo {author} {\bibfnamefont {T.~W.}\ \bibnamefont {Neely}},
  \bibinfo {author} {\bibfnamefont {D.~R.}\ \bibnamefont {Scherer}}, \bibinfo
  {author} {\bibfnamefont {A.~S.}\ \bibnamefont {Bradley}}, \bibinfo {author}
  {\bibfnamefont {M.~J.}\ \bibnamefont {Davis}}, \ and\ \bibinfo {author}
  {\bibfnamefont {B.~P.}\ \bibnamefont {Anderson}},\ }\href@noop {} {\bibfield
  {journal} {\bibinfo  {journal} {Nature}\ }\textbf {\bibinfo {volume} {455}},\
  \bibinfo {pages} {948} (\bibinfo {year} {2008})}\BibitemShut {NoStop}%
\bibitem [{\citenamefont {Sun}\ \emph {et~al.}(2017)\citenamefont {Sun},
  \citenamefont {Wen}, \citenamefont {Yoon}, \citenamefont {Liu}, \citenamefont
  {Steger}, \citenamefont {Pfeiffer}, \citenamefont {West}, \citenamefont
  {Snoke},\ and\ \citenamefont {Nelson}}]{sun2017bose}%
  \BibitemOpen
  \bibfield  {author} {\bibinfo {author} {\bibfnamefont {Y.}~\bibnamefont
  {Sun}}, \bibinfo {author} {\bibfnamefont {P.}~\bibnamefont {Wen}}, \bibinfo
  {author} {\bibfnamefont {Y.}~\bibnamefont {Yoon}}, \bibinfo {author}
  {\bibfnamefont {G.}~\bibnamefont {Liu}}, \bibinfo {author} {\bibfnamefont
  {M.}~\bibnamefont {Steger}}, \bibinfo {author} {\bibfnamefont {L.~N.}\
  \bibnamefont {Pfeiffer}}, \bibinfo {author} {\bibfnamefont {K.}~\bibnamefont
  {West}}, \bibinfo {author} {\bibfnamefont {D.~W.}\ \bibnamefont {Snoke}}, \
  and\ \bibinfo {author} {\bibfnamefont {K.~A.}\ \bibnamefont {Nelson}},\
  }\href@noop {} {\bibfield  {journal} {\bibinfo  {journal} {Physical review
  letters}\ }\textbf {\bibinfo {volume} {118}},\ \bibinfo {pages} {016602}
  (\bibinfo {year} {2017})}\BibitemShut {NoStop}%
\bibitem [{\citenamefont {Das}\ \emph {et~al.}(2012)\citenamefont {Das},
  \citenamefont {Sabbatini},\ and\ \citenamefont {Zurek}}]{das2012winding}%
  \BibitemOpen
  \bibfield  {author} {\bibinfo {author} {\bibfnamefont {A.}~\bibnamefont
  {Das}}, \bibinfo {author} {\bibfnamefont {J.}~\bibnamefont {Sabbatini}}, \
  and\ \bibinfo {author} {\bibfnamefont {W.~H.}\ \bibnamefont {Zurek}},\
  }\href@noop {} {\bibfield  {journal} {\bibinfo  {journal} {Scientific
  reports}\ }\textbf {\bibinfo {volume} {2}},\ \bibinfo {pages} {1} (\bibinfo
  {year} {2012})}\BibitemShut {NoStop}%
\bibitem [{\citenamefont {Berezinskii}(1972)}]{berezinskii1972destruction}%
  \BibitemOpen
  \bibfield  {author} {\bibinfo {author} {\bibfnamefont {V.}~\bibnamefont
  {Berezinskii}},\ }\href@noop {} {\bibfield  {journal} {\bibinfo  {journal}
  {Sov. Phys. JETP}\ }\textbf {\bibinfo {volume} {34}},\ \bibinfo {pages} {610}
  (\bibinfo {year} {1972})}\BibitemShut {NoStop}%
\bibitem [{\citenamefont {Kosterlitz}\ and\ \citenamefont
  {Thouless}(1973)}]{kosterlitz1973ordering}%
  \BibitemOpen
  \bibfield  {author} {\bibinfo {author} {\bibfnamefont {J.~M.}\ \bibnamefont
  {Kosterlitz}}\ and\ \bibinfo {author} {\bibfnamefont {D.~J.}\ \bibnamefont
  {Thouless}},\ }\href@noop {} {\bibfield  {journal} {\bibinfo  {journal}
  {Journal of Physics C: Solid State Physics}\ }\textbf {\bibinfo {volume}
  {6}},\ \bibinfo {pages} {1181} (\bibinfo {year} {1973})}\BibitemShut
  {NoStop}%
\bibitem [{\citenamefont {Bauer}\ \emph {et~al.}(2009)\citenamefont {Bauer},
  \citenamefont {Lettner}, \citenamefont {Vo}, \citenamefont {Rempe},\ and\
  \citenamefont {D{\"u}rr}}]{bauer2009control}%
  \BibitemOpen
  \bibfield  {author} {\bibinfo {author} {\bibfnamefont {D.~M.}\ \bibnamefont
  {Bauer}}, \bibinfo {author} {\bibfnamefont {M.}~\bibnamefont {Lettner}},
  \bibinfo {author} {\bibfnamefont {C.}~\bibnamefont {Vo}}, \bibinfo {author}
  {\bibfnamefont {G.}~\bibnamefont {Rempe}}, \ and\ \bibinfo {author}
  {\bibfnamefont {S.}~\bibnamefont {D{\"u}rr}},\ }\href@noop {} {\bibfield
  {journal} {\bibinfo  {journal} {Nature Physics}\ }\textbf {\bibinfo {volume}
  {5}},\ \bibinfo {pages} {339} (\bibinfo {year} {2009})}\BibitemShut {NoStop}%
\bibitem [{\citenamefont {Groszek}\ \emph {et~al.}(2018)\citenamefont
  {Groszek}, \citenamefont {Paganin}, \citenamefont {Helmerson},\ and\
  \citenamefont {Simula}}]{Groszek_2018}%
  \BibitemOpen
  \bibfield  {author} {\bibinfo {author} {\bibfnamefont {A.~J.}\ \bibnamefont
  {Groszek}}, \bibinfo {author} {\bibfnamefont {D.~M.}\ \bibnamefont
  {Paganin}}, \bibinfo {author} {\bibfnamefont {K.}~\bibnamefont {Helmerson}},
  \ and\ \bibinfo {author} {\bibfnamefont {T.~P.}\ \bibnamefont {Simula}},\
  }\href {\doibase 10.1103/physreva.97.023617} {\bibfield  {journal} {\bibinfo
  {journal} {Physical Review A}\ }\textbf {\bibinfo {volume} {97}} (\bibinfo
  {year} {2018}),\ 10.1103/physreva.97.023617}\BibitemShut {NoStop}%
\bibitem [{\citenamefont {Berloff}\ and\ \citenamefont
  {Youd}(2007)}]{berloff2007dissipative}%
  \BibitemOpen
  \bibfield  {author} {\bibinfo {author} {\bibfnamefont {N.~G.}\ \bibnamefont
  {Berloff}}\ and\ \bibinfo {author} {\bibfnamefont {A.~J.}\ \bibnamefont
  {Youd}},\ }\href@noop {} {\bibfield  {journal} {\bibinfo  {journal} {Physical
  review letters}\ }\textbf {\bibinfo {volume} {99}},\ \bibinfo {pages}
  {145301} (\bibinfo {year} {2007})}\BibitemShut {NoStop}%
\bibitem [{\citenamefont {Kalinin}\ and\ \citenamefont
  {Berloff}(2018)}]{kalinin2018networks}%
  \BibitemOpen
  \bibfield  {author} {\bibinfo {author} {\bibfnamefont {K.~P.}\ \bibnamefont
  {Kalinin}}\ and\ \bibinfo {author} {\bibfnamefont {N.~G.}\ \bibnamefont
  {Berloff}},\ }\href@noop {} {\bibfield  {journal} {\bibinfo  {journal} {New
  Journal of Physics}\ }\textbf {\bibinfo {volume} {20}},\ \bibinfo {pages}
  {113023} (\bibinfo {year} {2018})}\BibitemShut {NoStop}%
\bibitem [{\citenamefont {Berloff}(2004)}]{berloff2004pade}%
  \BibitemOpen
  \bibfield  {author} {\bibinfo {author} {\bibfnamefont {N.~G.}\ \bibnamefont
  {Berloff}},\ }\href@noop {} {\bibfield  {journal} {\bibinfo  {journal}
  {Journal of Physics A: Mathematical and General}\ }\textbf {\bibinfo {volume}
  {37}},\ \bibinfo {pages} {1617} (\bibinfo {year} {2004})}\BibitemShut
  {NoStop}%
\bibitem [{\citenamefont {White}\ \emph {et~al.}(2014)\citenamefont {White},
  \citenamefont {Anderson},\ and\ \citenamefont {Bagnato}}]{white2014vortices}%
  \BibitemOpen
  \bibfield  {author} {\bibinfo {author} {\bibfnamefont {A.~C.}\ \bibnamefont
  {White}}, \bibinfo {author} {\bibfnamefont {B.~P.}\ \bibnamefont {Anderson}},
  \ and\ \bibinfo {author} {\bibfnamefont {V.~S.}\ \bibnamefont {Bagnato}},\
  }\href@noop {} {\bibfield  {journal} {\bibinfo  {journal} {Proceedings of the
  National Academy of Sciences}\ }\textbf {\bibinfo {volume} {111}},\ \bibinfo
  {pages} {4719} (\bibinfo {year} {2014})}\BibitemShut {NoStop}%
\end{thebibliography}%



\appendix*

\end{document}